\documentclass[prd,twocolumn,superscriptaddress,
preprintnumbers,nofootinbib,showpacs]{revtex4-1}
\usepackage{graphicx}
\usepackage{amsmath}
\usepackage{amssymb}
\usepackage{color}
\usepackage{hyperref}
\usepackage{url}
\usepackage{breakurl}
 
\newcommand{\be}{\begin{equation}}
\newcommand{\ee}{\end{equation}}
\newcommand{\bea}{\begin{eqnarray}}
\newcommand{\eea}{\end{eqnarray}}

\def\met{\slash{\!\!\!\!E}_{\text{T}}}
\def \red#1 {\textcolor{red}{#1}\ }

\begin{document}

\title{Higgs-flavon mixing and LHC phenomenology \\in a simplified model of broken flavor symmetry}

\author{Edmond L. Berger}
\email{berger@anl.gov}
\affiliation{High Energy Physics Division, Argonne National Laboratory, 
Argonne, IL 60439, USA}

\author{Steven B. Giddings}
\email{giddings@physics.ucsb.edu}
\affiliation{Department of Physics, University of California, 
Santa Barbara, CA 93106, USA}

\author{Haichen Wang}
\email{haichenwang@lbl.gov}
\affiliation{Lawrence Berkeley National Laboratory, Berkeley, CA 94720, USA}

\author{Hao Zhang}
\email{zhanghao@physics.ucsb.edu}
\affiliation{Department of Physics, University of California, 
Santa Barbara, CA 93106, USA}
\affiliation{Center for High Energy Physics, Peking University, Beijing 100871, China}

\begin{abstract}
The LHC phenomenology of a low-scale gauged flavor symmetry 
model with inverted hierarchy is studied, through introduction of a simplified model of broken flavor symmetry. A 
new scalar (a flavon) and a new neutral top-philic 
massive gauge boson emerge with mass in the TeV range along 
with a new heavy fermion associated 
with the standard model top quark.  After checking constraints from electroweak 
precision observables,  we investigate the influence of the model 
on Higgs boson physics,  notably  on its 
production cross section and decay branching fractions.  Limits  
on the flavon $\varphi$ from heavy Higgs boson 
searches at the LHC at 7 and 8 TeV are presented.  
The branching fractions of the flavon are computed 
as a function of the flavon mass and the Higgs-flavon mixing angle.  
We also explore possible discovery 
of the flavon at 14 TeV, particularly via the $\varphi \rightarrow Z^0Z^0$ decay channel 
in the $2\ell2\ell'$ final state, and through standard model 
Higgs boson pair production $\varphi \rightarrow hh$ in the $b\bar{b}\gamma\gamma$ 
final state.   We conclude that 
the flavon mass range up to $500$ GeV could probed down to 
quite small values of the Higgs-flavon 
mixing angle with 100 fb$^{-1}$ of integrated luminosity at 14 TeV. 
\end{abstract}


\maketitle

\section{Introduction}
\label{sec:intro}

The standard model (SM) of particle physics describes physics at the electroweak symmetry 
breaking (EWSB) scale of the visible sector remarkably well.  With the discovery of a
Higgs boson behaving much like that of the SM
at the  Large Hadron Collider (LHC) \cite{Aad:2012tfa,Chatrchyan:2012ufa}, 
all of the expected SM particles have been detected.  Attention has turned to precise 
determination of the properties of the Higgs boson, notably its decay branching fractions, and 
to the search for possible new physics beyond the SM.  Given current precision, the branching 
fractions allow some, if limited, deviations from SM predictions.   

The SM poses several puzzles.  These include the origin of the fermion 
mass hierarchy and the flavor structure parametrized in the well-known CKM matrix \cite{Cabibbo:1963yz,Kobayashi:1973fv}.  The 
dynamics of flavor mixing is well described within a framework of three generations of quarks.  
The fact that no significant deviations from SM predictions have appeared in any flavor-related physics processes indicates that 
that any TeV scale new physics (NP) does not introduce any important new source of flavor change 
or CP violation beyond the SM. 
This hints at flavor symmetry (horizontal symmetry) in a 
NP model. The idea that the NP interactions are invariant under a flavor symmetry group is known 
as minimal flavor violation (MFV)~\cite{D'Ambrosio:2002ex}.  

In the MFV scenario, the SM flavor symmetry is broken explicitly by the non-vanishing SM
Yukawa coupling constants. This symmetry could nevertheless be a true symmetry of nature at some high
energy scale but broken by non-zero vacuum expectation values (vev's) of scalar fields which
are usually called flavons. In such a case, the SM Yukawa coupling constants are related
to the ratio between the vev's of flavons and some cutoff scale. 

If the full non-abelian flavor symmetry of the fermion kinetic terms and gauge couplings is a global symmetry broken by flavons, 
this yields Goldstone bosons subject to stringent constraints.  This problem can be avoided if the non-abelian flavor symmetry is gauged, giving mass to the Goldstone modes.\footnote{Other scenarios include abelian flavor symmetries \cite{Leurer:1992wg,Leurer:1993gy} 
and discrete flavor symmetries\cite{Zwicky:2009vt}; these are not considered in this
work.}   
Anomaly cancellation in such a theory requires
introduction of exotic fermions\cite{Berezhiani:1983rk,Berezhiani:1983hm,
Berezhiani:1990wn,Grinstein:2010ve,Feldmann:2010yp,Krnjaic:2012aj}
to cancel gauge anomalies.   These mix with the SM fermions.  The SM
fermion masses are the smaller eigenvalues of the mass matrix and are proportional to the inverse
of the flavon vev's, corresponding to an inverted hierarchy. The masses of all NP particles, 
such as the extra fermions, flavons, and flavor gauge bosons,
are controlled by the flavon vev's and therefore are approximately proportional to the inverse of the SM Yukawa
constants. Constraints from low energy precision observables and flavor physics are carefully considered in \cite{Grinstein:2010ve,Buras:2011wi}.

The lightest new particles in such a  flavor symmetry model with inverted hierarchy are the exotic fermions, 
the flavon which couples to the third generation of SM fermions, and a massive top-philic gauge boson.  
Their masses could be at the TeV scale, and it should be possible to search for them at the 
LHC.  The LHC phenomenology of the flavon, the top-philic gauge boson 
and the heavy fermion partner of the top-quark might be 
interestingly rich.  

In this paper we do not focus on details of flavor physics {\em per se}.  Rather, we address the implications 
of flavons and the heavy fermion partner of the top-quark for Higgs boson physics, and the LHC phenomenology 
of a simplified flavor symmetry model with inverted hierarchy.  We begin in Sec. \ref{sec:model} with an 
explanation of the motivation and origin for the inverted hierarchy in a flavor symmetry model. The simplified
Lagrangian and the mass eigenstates are shown in this section also.
In Sec. \ref{sec:ewpo}, we briefly review the constraints from electroweak 
precision observables (EWPO) and flavor violation experiments.
We study the effects of the flavor symmetry model on the production and decay properties of the 
SM Higgs boson in Sec. \ref{sec:higgs}.   The inclusive Higgs boson production cross section is 
suppressed relative to the SM by a factor $c_H^2 = \cos^2 \theta_H$, where $\theta_H$ is the mixing angle of 
the scalar flavon and the Higgs boson.  This suppression is allowed by the LHC data at 7 and 
8 TeV, within limits.  We show that most of the Higgs boson couplings to SM particles are 
just rescaled by a factor $c_H$, including the loop induced $hgg$ and $h\gamma\gamma$ vertices 
in the heavy fermion limit.  The $hZ^0\gamma$ vertex deviates from the simple $c_H$ rescaling, but the 
deviation is not huge.   The Higgs boson decay branching ratios are nearly unchanged relative to the SM 
since every sizable partial width is changed by an overall factor $c_H^2$.   In Sec. \ref{sec:lhc}, we 
investigate limits on the flavon from LHC data at 7 and 8 TeV and possible signals of the flavon at 14 TeV.  
Flavon searches at the LHC can focus on the SM Higgs-like decay channels ($Z^0Z^0$, $W^+W^-$) and on 
the Higgs boson pair decay channel $\varphi
\rightarrow hh$.  We compute and display the decay branching 
fractions of the flavon as a function of the flavon mass and mixing angle $\theta_H$.  
Only the $W^+W^-, Z^0Z^0, hh$ and $t\bar t$ channels are significant in flavon decay.
We examine bounds on the parameter space of flavons from heavy Higgs boson searches 
at 7 and 8 TeV. Because the flavon can be produced singly, if it decays into the 
$hh$ final state with an appreciable decay branching ratio, the Higgs pair cross section 
will be enhanced significantly by this resonance effect.
We perform a detailed simulation of the signal and backgrounds for the 
$\varphi\rightarrow hh \rightarrow b \bar{b} 
\gamma \gamma$ channel at 14 TeV for an assumed integrated 
luminosity of $100 \rm{fb}^{-1}$, deriving both $2$ standard deviation exclusion limits and $5$ standard 
deviation observation bounds as a function of flavon mass.  In some regions of parameter space the 
search for a $hh$ signal will give a stronger constraint on the NP model than the $Z^0Z^0$ channel.    
Our conclusions are summarized in Sec. \ref{sec:con}.  

\section{From gauged flavor symmetry to a simplified model of broken flavor symmetry}
\label{sec:model}

The flavor symmetry of the quark kinetic terms and gauge couplings is 

\be \label{FlavSym}
G_f=
U(3)_{Q_L}\otimes U(3)_{U_R}\otimes U(3)_{D_R}\ ,
\ee
If this symmetry is gauged with only SM fermions present, the theory is anomalous.  
A ``minimal" model of new fermions that cancel the anomalies was described in  \cite{Grinstein:2010ve}.  This model has exotic fermion partners of the 
SM quarks, flavor gauge bosons, and two scalar flavon fields $Y_u$ and $Y_d$ for the up-like and down-like quarks. $U(1)_{Q_L}$ remains anomalous, but the rest of the flavor symmetry \eqref{FlavSym} is taken to be gauged.
The most general renormalizable interaction Lagrangian between the flavon fields and the SM and exotic fermions takes the form 
\bea
\mathcal{L}_{\text{UV}}&=&\mathcal{L}_{\text{kinetic\ +\ gauge}}\nonumber\\
&&-(-\lambda_u\bar Q_L \tilde{H} \Psi_{uR}+\lambda_u^\prime\bar \Psi_u Y_u \Psi_{uR}
+M_u\bar\Psi_{u}U_R\nonumber\\
&&-\lambda_d\bar Q_L H \Psi_{dR}+\lambda_d^\prime\bar \Psi_d Y_d \Psi_{dR}
+M_d\bar\Psi_{d}D_R+{\text{h.c.}})\nonumber\\
&&-V(Y_u,Y_d,H)\  .
\label{eq:opeuv}
\eea
Here $Q_L$, $U_R$, $D_R$ are the SM quark fields,  
$\Psi_{u}$, $\Psi_{uR}$, $\Psi_{d}$, and  $\Psi_{dR}$ are the partner fermion fields, $H$ is the 
SM Higgs doublet field, and $\tilde H_i\equiv \varepsilon_{ij}H_j$ where {$\varepsilon_{ij}$ is the anti-symmetric 
tensor with $\varepsilon_{12}=1$.  $\lambda$ and $\lambda^\prime$ are dimensionless parameters,  
$M$ is a parameter with the dimensions of mass, and $V$ is the scalar potential.   The representations
under the gauge groups to which these these fields belong is shown in TABLE \ref{tab:repuv}. 
One can verify that both the SM gauge symmetry and 
the flavor gauge symmetry are anomaly free with the 
contributions from the exotic fermion fields.
If flavor symmetry breaks via flavon vevs with $\langle Y\rangle \gg M$, the masses of the SM fermions are inversely proportional 
to the vev of the corresponding flavon field component, resulting in the asserted inverted hierarchy.
\begin{table}[htdp]
\caption{The representation of the fields in Eq (\ref{eq:opeuv}) 
under the SM gauge group and the flavor symmetry group.}
\begin{center}
\begin{tabular}{c|c|c|c|c|c|c}
\hline       &$SU(3)_{Q_L}$&$SU(3)_{U_R}$&$SU(3)_{D_R}$&$SU(3)_c$&$SU(2)_L$&$U(1)_Y$ \\\hline
\hline $Q_L$ & {\bf 3} & {\bf 1} & {\bf 1}& {\bf 3} & {\bf 2} & $1/6$    \\
\hline $U_R$ & {\bf 1} & {\bf 3} & {\bf 1}& {\bf 3} & {\bf 1} & $2/3$    \\
\hline $D_R$ & {\bf 1} & {\bf 1} & {\bf 3} & {\bf 3} & {\bf 1} & -$1/3$   \\
\hline $\Psi_{u}$ & {\bf 1} & {\bf 3} & {\bf 1}& {\bf 3} & {\bf 1} & $2/3$    \\
\hline $\Psi_{d}$ & {\bf 1} & {\bf 1} & {\bf 3}& {\bf 3} & {\bf 1} & -$1/3$    \\
\hline $\Psi_{uR}$ & {\bf 3} & {\bf 1} & {\bf 1}& {\bf 3} & {\bf 1} & $2/3$    \\
\hline $\Psi_{dR}$ & {\bf 3} & {\bf 1} & {\bf 1}& {\bf 3} & {\bf 1} & -$1/3$    \\
\hline $Y_u$ & {$\bf \bar3$} & {\bf 3} & {\bf 1}& {\bf 1} & {\bf 1} & 0    \\
\hline $Y_d$ & {$\bf \bar3$} & {\bf 1} & {\bf 3}& {\bf 1} & {\bf 1} & 0    \\
\hline $H$ & {\bf 1} & {\bf 1} & {\bf 1} & {\bf 1} & {\bf 2} & $1/2$  \\
\hline
\end{tabular}
\end{center}
\label{tab:repuv}
\end{table}%

The large hierarchy between the masses of the SM quarks thus corresponds to a large hierarchy
between the vevs of the flavons which suggests 
that the flavor symmetry could be broken sequentially \cite{Feldmann:2009dc}.
Guided by this realization, we assume that 
the breaking of the Lagrangian in Eq (\ref{eq:opeuv}) occurs in a sequence of steps. 
From the effective field theory point of view, one successively integrates out heavy degrees of freedom.

If we integrate out the heavy  degrees of freedom associated with the first and second generations, we are left with a simplified flavor symmetry model with a manageable 
number of BSM degrees of freedom (a flavon, an exotic fermion, and a massive vector boson) 
associated with the top-quark 
and bottom-quark sectors. 
However, because the vev of the flavon associated with the bottom-quark is nearly two orders of 
magnitude larger than the vev of the flavon associated with the top-quark, this suggests finally integrating out the flavon associated with the bottom-quark.
Thus, at the TeV scale, we have the effective Lagrangian
\be
\mathcal{L}_{\text{topflavor}}=\lambda\bar Q_L \tilde{H} \Psi_{tR}-\lambda^\prime\bar \Psi_{t} \Phi 
\Psi_{tR}-M\bar \Psi_{t} t_R+{\text{h.c.}},
\label{eq:ope}
\ee
where  $\Phi$ is a complex flavon associated with the top-quark. There is also a residual 
$U(1)$ gauged flavor symmetry under which only the $\Phi$, $\Psi_t$ and $t_R$ fields carry (the same) charge.
Chiral phase rotations of the fermions allow us to take, 
without loss of generality, $\lambda,\lambda',M>0$. We can consider the Lagrangian
\eqref{eq:ope} separately from our discussion of the higher-scale flavor structure, as 
a simplified model extending the top and Higgs sectors. The usual SM Yukawa interactions for the remaining 
fermions are added to this Lagrangian.\footnote{There are other models with 
similar Lagrangians, although arising from different motivations. For example,
see \cite{Xiao:2014kba,He:2014ora}.}

After EWSB and FSB, 
\bea
H&=&\left(\begin{array}{c}0 \\\frac{v+\tilde h}{\sqrt 2}\end{array}\right),\\
\Phi&=&\frac{\tilde \varphi+v_\varphi}{\sqrt2},
\eea
in the unitary gauge, where $v=246$ GeV is the vev of the Higgs field,
$\tilde h$ is the physical degree of freedom of the SM Higgs doublet field, and $\tilde \varphi$
is the physical degree of freedom of the top flavon. 
The mass eigenstates are linear 
combinations of $\tilde h$ and $\tilde \varphi$ which will be given below.

The mass matrix of the fermions is
\bea
\mathcal{L}_M=-\left(\begin{array}{cc}\bar t_L & \bar \Psi_{t}\end{array}\right)
\left(\begin{array}{cc}0 & -\lambda v/\sqrt2 \\ M & \lambda^\prime v_\varphi/\sqrt2\end{array}\right)
\left(\begin{array}{c}t_R \\ \Psi_{tR}\end{array}\right).
\eea
It can be diagonalized by separate left and right rotations.  This results in mass eigenvalues given by
\bea
m_t^2&=&\frac{1}{4}\biggl[-\sqrt{\left(2M^2+\lambda^2v^2+\lambda^{\prime2}v_\varphi^2\right)^2-8\lambda^2M^2v^2}\nonumber\\
&&+2M^2+\lambda^2v^2+\lambda^{\prime2}v_\varphi^2\biggr],\nonumber\\
m_T^2&=&\frac{1}{4}\biggl[\sqrt{\left(2M^2+\lambda^2v^2+\lambda^{\prime2}v_\varphi^2\right)^2-8\lambda^2M^2v^2}\nonumber\\
&&+2M^2+\lambda^2v^2+\lambda^{\prime2}v_\varphi^2\biggr]\ ,
\label{eq:mass}
\eea
which are positive and real for real $\lambda,\lambda',M$.
Here $m_t$ should be the running mass of the top quark which is 163 GeV \cite{Alekhin:2012py}.
One can find
$\lambda^\prime v_\varphi$ in terms of the other parameters, if $m_t$ is fixed to be the mass of the SM top quark.  
Reality of $\lambda^\prime v_\varphi$ requires
\be
\left(M^2-m_t^2\right)\left(\lambda^2v^2-2m_t^2\right)>0.
\ee
The case $m_t>M$ and $\lambda v<\sqrt2 m_t$ would correspond to the SM top quark being the heavier fermion; 
we do not treat this scenario
because a light colored fermion $T$ which also couples to the electroweak 
gauge boson and the Higgs boson would be highly constrained by current data. 
Thus, $T$ will denote the heavy partner of the top quark.

When $M>m_t$ and $\lambda v>\sqrt2 m_t$, the other mass eigenvalue is
\be
m_T=\frac{Mv\lambda}{\sqrt2 m_t}.
\ee
The right and left components of the two fermion mass eigenstates, 
the SM top quark $t$ and a heavy fermion $T$, are 
\bea
\left(\begin{array}{c}t_R \\ \Psi_{tR}\end{array}\right)&=&P_R
\left(\begin{array}{cc}\cos\theta_R & \sin\theta_R \\ -\sin\theta_R &\cos\theta_R\end{array}\right)
\left(\begin{array}{c}t \\T\end{array}\right),\\
\left(\begin{array}{c}t_L \\ \Psi_{t}\end{array}\right)&=&P_L
\left(\begin{array}{cc}\cos\theta_L & \sin\theta_L \\ -\sin\theta_L &\cos\theta_L\end{array}\right)
\left(\begin{array}{c}t \\T\end{array}\right).
\eea
It is easy to derive
\bea
s_L\equiv \sin\theta_L &=&-\frac{m_t\sqrt{\lambda^2v^2-2m_t^2}}{\sqrt{M^2v^2\lambda^2-2m_t^4}},\\
c_L\equiv \cos\theta_L&=&\frac{\lambda v\sqrt{M^2-m_t^2}}{\sqrt{M^2v^2\lambda^2-2m_t^4}},\\
s_R\equiv \sin\theta_R &=&\frac{\sqrt2m_t\sqrt{M^2-m_t^2}}{\sqrt{M^2v^2\lambda^2-2m_t^4}},\\
c_R\equiv \cos\theta_R &=&\frac{M\sqrt{\lambda^2v^2-2m_t^2}}{\sqrt{M^2v^2\lambda^2-2m_t^4}}.
\eea
The Yukawa interactions are therefore
\bea
\mathcal{L}_{Y}&=&\frac{\lambda \tilde h}{\sqrt2}\biggl[-\bar ttc_Ls_R
+\bar t\left(-s_Ls_RP_L+c_Lc_RP_R\right)T\nonumber\\
&&+\bar T\left(c_Lc_RP_L-s_Ls_RP_R\right)t+
\bar TTs_Lc_R\biggr]\nonumber\\
&&-\frac{\lambda^\prime \tilde \varphi}{\sqrt2}
\biggl[\bar tts_Ls_R-\bar t\left(c_Ls_RP_L+s_Lc_RP_R\right)T\nonumber\\
&&-\bar T\left(s_Lc_RP_L+c_Ls_RP_R\right)t+
\bar TTc_Lc_R\biggr],
\eea
while the gauge interactions are
\bea
\mathcal{L}_{K}&=&\frac{2es_W}{3c_W}\bar T\gamma^\mu
\left[\left(1-\frac{3s_L^2}{4s_W^2}\right)P_L
+P_R\right]TZ_\mu\nonumber\\
&&+\frac{2es_W}{3c_W}
\bar t\gamma^\mu\left[\left(1-\frac{3c_L^2}{4s_W^2}\right)P_L
+P_R\right]tZ_\mu\nonumber\\
&&-\frac{es_Lc_L}{2s_Wc_W}\left(\bar t \gamma^\mu P_LT
+\bar T \gamma^\mu P_Lt\right)Z^\mu\nonumber\\
&&-\frac{ec_L}{\sqrt2s_W}\bar t\gamma^\mu P_LbW^+_\mu
-\frac{ec_L}{\sqrt2s_W}\bar b\gamma^\mu P_LtW^-_\mu
\nonumber\\
&&-\frac{es_L}{\sqrt2s_W}\bar T\gamma^\mu P_LbW^+_\mu
-\frac{es_L}{\sqrt2s_W}\bar b\gamma^\mu P_LTW^-_\mu
\nonumber\\
&&-\frac{2e}{3}\bar t\gamma^\mu tA_\mu-\frac{2e}{3}\bar T\gamma^\mu TA_\mu,
\eea
where 
$s_W$ ($c_W$) is the sine (cosine) 
of the weak angle.

In the scalar potential, the trilinear interaction 
between $\Phi$ and $H$ is forbidden by the 
$SU(2)_L$ and flavor symmetries.  However, the interaction 
\be
(\Phi^*\Phi)(H^\dagger H)
\label{eq:mix}
\ee
is still allowed. This term can be generated through a combined top-quark and heavy fermion $T$ loop
in the one-loop effective potential even it is forbidden artificially at tree-level 
(FIG. \ref{fig:eff_potential}).   
\begin{figure}[!htb]
\includegraphics[scale=0.35,clip]{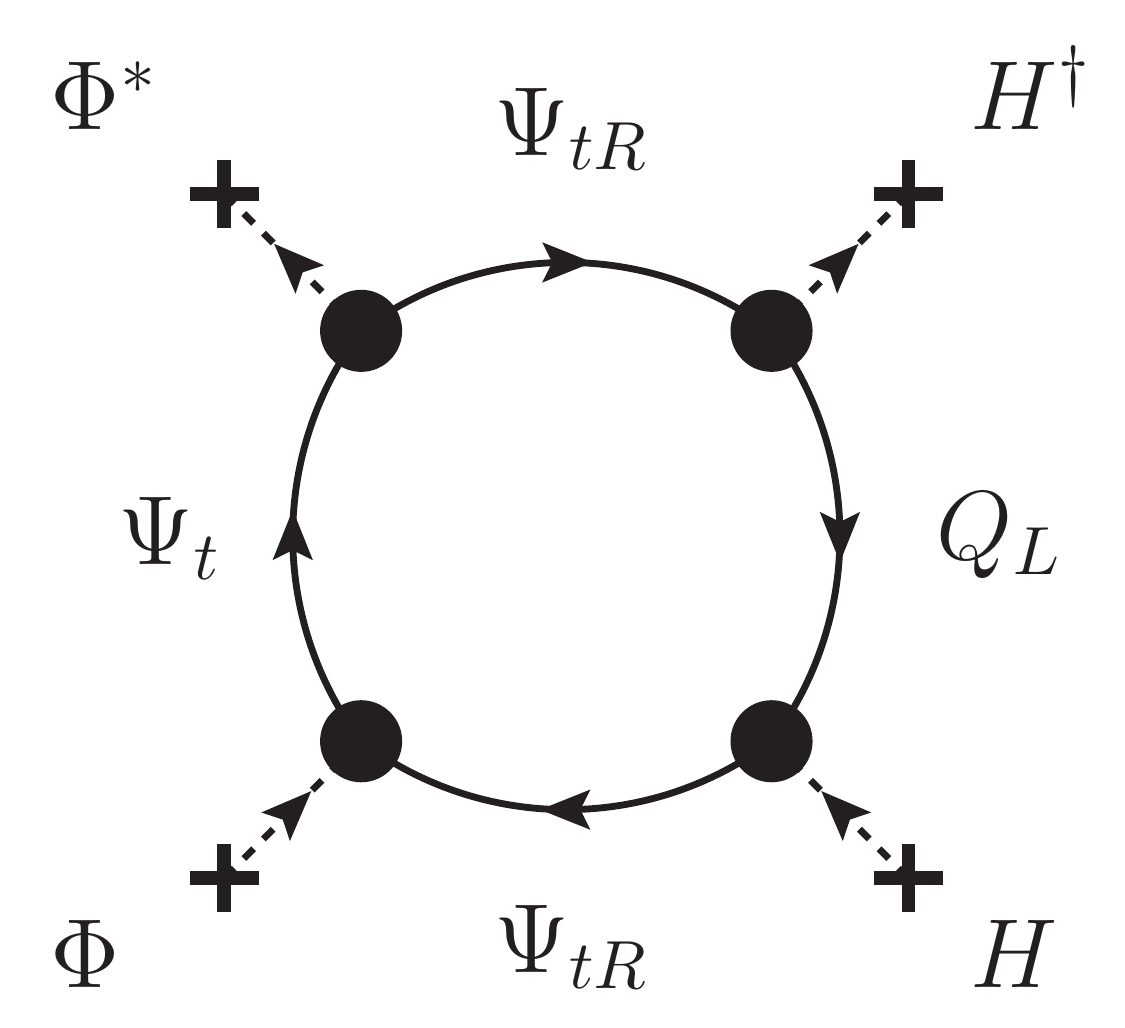}
\caption{Contribution to the one-loop effective potential from Eq (\ref{eq:ope}). 
Each black dot means an insertion of the vertex and a zero-momentum 
external scalar.
\label{fig:eff_potential} }
\end{figure}
The general renormalizable Lagrangian of the scalar fields of the complex gauge singlet extension 
of the SM can be written as
\be
\mathcal{L}=\left(D_\mu H\right)^\dagger \left(D^\mu H\right)+
\left(D_\mu \Phi\right)^*\left(D^\mu \Phi\right)-V\left(H,\Phi\right)
\label{eq:lag_scalar}
\ee
with scalar potential
\bea
V\left(H,\Phi\right)&=&
\frac{\lambda_H}{2}\left(H^\dagger H\right)^2+\frac{\lambda_\Phi}{2}\left(\Phi^*\Phi\right)^2
-\frac{1}{2}m_H^2\left(H^\dagger H\right)\nonumber\\
&&-\frac{1}{2}m_\Phi^2(\Phi^*\Phi)+\xi (\Phi^*\Phi)\left(H^\dagger H\right)\ ;
\eea
the parameters $\lambda_H$ and $\lambda_\Phi$ describe the self-interactions of the 
Higgs field and the flavon field. The gauge covariant derivative of the SM doublet 
Higgs field is the same as the one in the SM. The
gauge covariant derivative of the flavon is 
\be
D_\mu \Phi=\partial_\mu\Phi-ig_TZ_{T\mu}\Phi,
\ee
where $g_T$ is the gauge coupling constant of the residual $U(1)$ flavor gauge symmetry,
and $Z_{T\mu}$ is the gauge field of the residual $U(1)$ flavor gauge symmetry.

The vev's of neutral components are found to be
\bea
v&=&\sqrt{\frac{\lambda_\Phi m_H^2-\xi m_\Phi^2}{\lambda_H \lambda_\Phi-\xi^2}},\\
v_\varphi&=&\sqrt{\frac{\lambda_H m_\Phi^2-\xi m_H^2}{\lambda_H \lambda_\Phi-\xi^2}}.
\eea
Avoidance of a flat direction of the vacuum requires
\be
\lambda_H \lambda_\Phi-\xi^2>0.
\ee
The vev of the $SU(2)_L$ doublet field is determined by the weak interaction coupling 
constant and the masses of the SM massive gauge bosons.  Therefore there is a constraint
\be
\sqrt{\frac{\lambda_\Phi m_H^2-\xi m_\Phi^2}{\lambda_H \lambda_\Phi-\xi^2}}=246{\text{GeV}}.
\ee
The physical degree of freedom which is dominated 
by $\tilde{h}$ should be the SM-like 
Higgs boson $h$.   Its mass should be $m_h=125.4$ GeV \cite{CMS-PAS-HIG-13-005,Aad:2014aba}.
 Assuming the other scalar field mass eigenstate has mass $m_\varphi$, 
we can solve for the parameters $m_H$ and $\xi$, and present the results in 
terms of $m_h, v, m_\varphi, \lambda_H$ and $\lambda_\Phi$, where 
the first two parameters are determined by current experiments. We can also determine 
$\lambda^\prime$ using these parameters as
\be
\lambda^\prime=\frac{1}{m_t}\sqrt{\frac{\lambda_\Phi
\left(M^2-m_t^2\right)\left(\lambda^2v^2-2m_t^2\right)}
{m_\varphi^2+m_h^2-\lambda_H v^2}}.
\ee

We define the mass eigenstates $\left(h,\varphi\right)$ of the scalar fields  by
\be
\left(\begin{array}{c}\tilde h \\\tilde \varphi\end{array}\right)=
\left(\begin{array}{cc}\cos\theta_H & \sin\theta_H \\ 
-\sin\theta_H &\cos\theta_H\end{array}\right)
\left(\begin{array}{c}h \\\varphi\end{array}\right),
\ee
where the rotation angle $\theta_H$ is given by
\bea
\sin\theta_H&=&\sqrt{\frac{\lambda_H v^2-m_h^2}{m_\varphi^2-m_h^2}},\\
\cos\theta_H&=&\sqrt{\frac{m_\varphi^2-\lambda_H v^2}{m_\varphi^2-m_h^2}}.
\eea
The deviation of the Higgs field self-interaction strength $\lambda_H$
from its value $\lambda_H^{SM}=m_h^2/v^2$ in the SM can be written as 
\be
\lambda_H\equiv\lambda_H^{SM}+\frac{m_\varphi^2-m_h^2}{v^2}\sin^2\theta_H.
\ee

The additional massive gauge boson $Z_{T\mu}$, whose
mass is $m_{Z_T}=g_T\sqrt{\left(m_\varphi^2c_H^2+m_h^2
s_H^2\right)/\lambda_\Phi}$,  
couples at tree-level only to the SM top-quark, the heavy fermion, the flavon, and the Higgs boson 
 through
\bea
\mathcal{L}&=&g_TZ_{T\mu}\bigl[\bar t\gamma^\mu\left(s_L^2P_L+c_R^2P_R\right)t
+\bar T\gamma^\mu\left(c_L^2P_L+s_R^2P_R\right)T\nonumber\\
&-&
\bar t\gamma^\mu\left(s_Lc_LP_L-s_Rc_RP_R\right)T
-\bar T\gamma^\mu\left(s_Lc_LP_L-s_Rc_RP_R\right)t\bigr]\nonumber\\
&+&g_Tm_{Z_T}c_H\varphi Z_{T\mu}Z^\mu_T
-g_Tm_{Z_T}s_HhZ_{T\mu}Z^\mu_T\nonumber\\
&+&\frac{1}{2}g_T^2c_H^2\varphi^2Z_{T\mu}Z^\mu_T
+\frac{1}{2}g_T^2s_H^2h^2Z_{T\mu}Z^\mu_T\nonumber\\
&-&g_T^2s_Hc_Hh\varphi Z_{T\mu}Z^\mu_T.
\eea
Searching for such a top-philic gauge boson is a 
challenging task at colliders when it does not mix with the SM $Z^0$
at tree-level 
\cite{Chiang:2007sf,Chen:2008za,Jackson:2009kg,Hsieh:2010zr,Berger:2011xk}.

In summary, this section has presented a 
simplified model of spontaneous flavor symmetry breaking, which
arises from the 
the gauged flavor symmetry 
model with inverted hierarchy in which only degrees of freedom related to the 
third generation are considered. There is a heavy fermion $T$ which mixes
with the SM top-quark, a heavy scalar flavon $\varphi$ which mixes with the
SM-like Higgs boson, and a heavy top-philic vector boson $Z_{T\mu}$. 
The basic couplings that we will need have been presented in this section.

\section{Constraints from the electroweak precision observables}
\label{sec:ewpo}
The interactions between the fermions and the SM gauge bosons are different 
in the low-scale gauged flavor symmetry model from the SM interactions.  The 
new scalar also couples to the SM gauge bosons such that the strength of these 
interactions should be constrained by SM electroweak precision observables. 
The modification of the charged current will also change the prediction of  
$b\to s\gamma$.  Constraints from the EWPO and flavor physics were  
considered in the original paper \cite{Grinstein:2010ve}.   In this work, we 
rexamine the EWPO constraints and include the contribution from the flavon 
$\varphi$ in our calculation.

\begin{figure}[!htb]
\includegraphics[scale=0.35,clip]{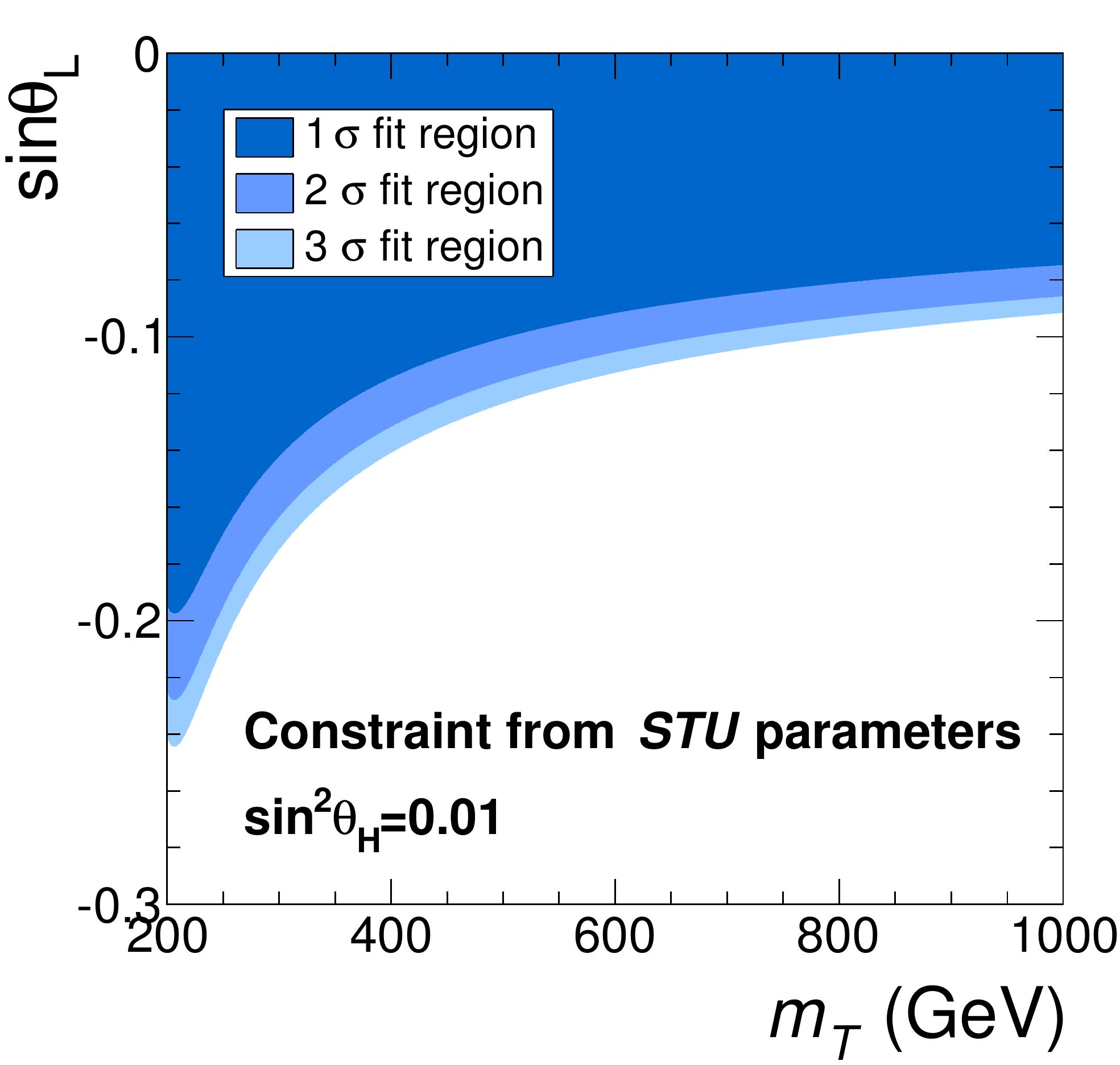}
\includegraphics[scale=0.35,clip]{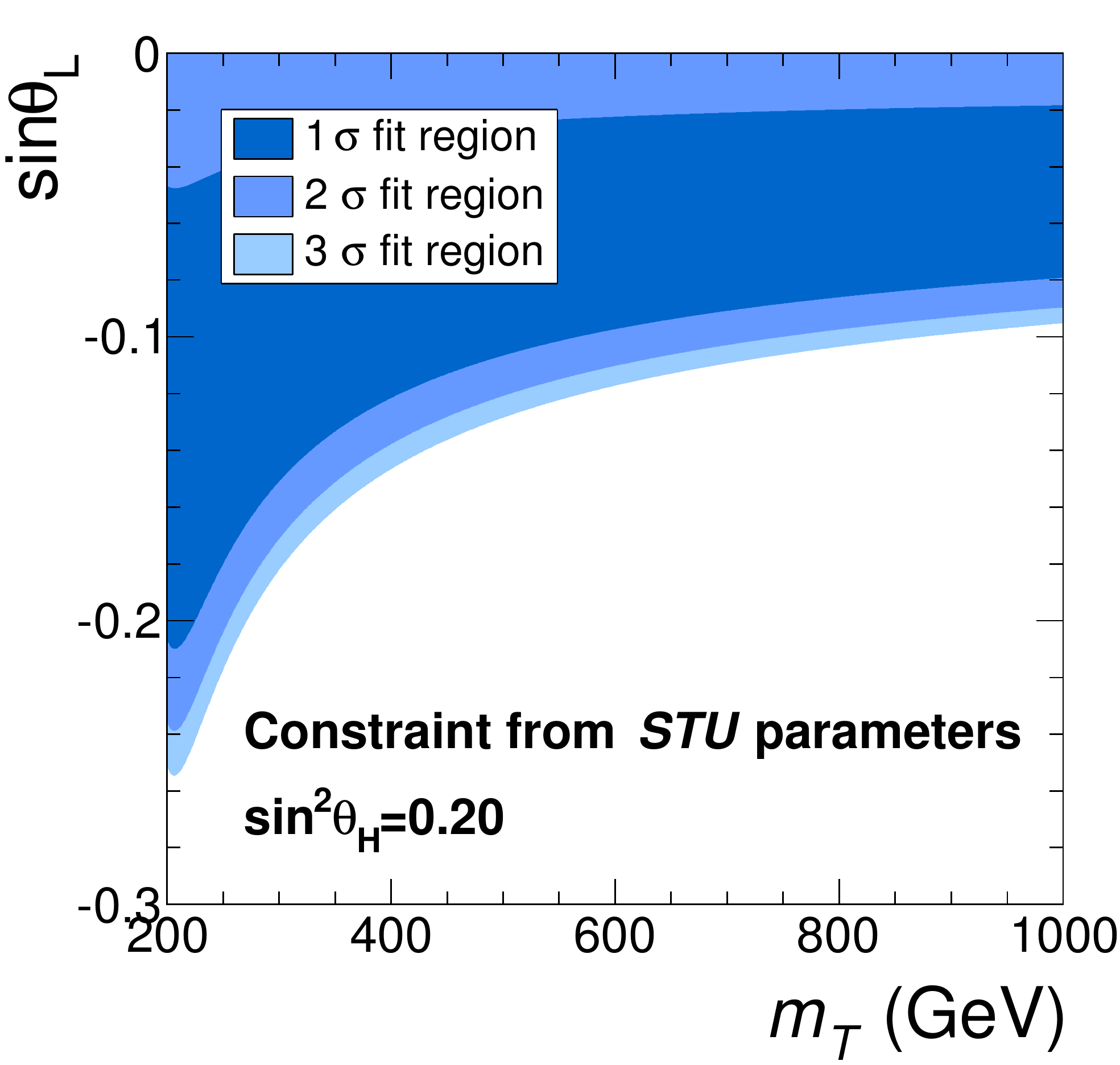}
\caption{The constraint on the mass of the heavy fermion $m_T$ and the 
mixing angle of the left handed fermions $s_L$ from 
the electroweak oblique parameters.  The flavon mass $m_\varphi$ is chosen to be 300 GeV.  
In the upper panel, the scalar mixing angle is $s_H^2=0.01$. 
In the lower panel, $s_H^2=0.20$.
\label{fig:expo} }
\end{figure}

Because the new physics effects occur above the $Z$-pole, their  
influence on the SM EWPO can be described with the well known oblique parameters 
$S$, $T$, $U$ \cite{Peskin:1990zt}.  When the SM reference values 
of $m_t$ and $m_h$ are chosen to be 
\be
m_{h,{\text{ref}}}=126{\text{GeV}}, ~~m_{t,{\text{ref}}}=173{\text{GeV}},
\ee
the best fit values of the oblique parameters are \cite{Baak:2012kk}
\bea
S&=&0.03\pm0.10,\nonumber\\
T&=&0.05\pm0.12,\\
U&=&0.03\pm0.10,\nonumber
\eea
while the correlation coefficients matrix is \cite{Baak:2012kk}
\bea
\left(\begin{array}{ccc}1 & 0.89 & -0.54 \\0.89 & 1 & -0.83 \\-0.54 & -0.83 & 1
\end{array}\right).
\eea

The contribution from the exotic real scalar boson $\varphi$ to the oblique 
parameters \cite{Barger:2007im} is suppressed by the
mixing angle $\theta_H$.
The contribution from the third generation is given in \cite{Grinstein:2010ve}. 
Using a $\chi^2$ check, we show the one standard deviation
(1$\sigma$), 2$\sigma$ and 3$\sigma$ fit regions in FIG. \ref{fig:expo}.
A detailed analysis of $\Delta F=2$ flavor physics observables and of
$B\to X_s\gamma$ in this model is presented in Ref. \cite{Buras:2011wi}.  After 
inclusion of the contributions from the gauge bosons of the flavor group,
this model could resolve the $\varepsilon_K-S_{\psi K_S}$ tension but 
result in a more serious tension from $\Delta M_{B_{d,s}}$ and 
$R_{BR/\Delta M}$ than in the SM.  Contributions from the flavons are not 
included.  Interested 
readers can find the constraints in Ref. \cite{Buras:2011wi}.

\section{Higgs Boson Physics}
\label{sec:higgs}

In this section, we investigate Higgs boson physics in  
the simplified flavor symmetry model under consideration.
The interactions between the SM-like Higgs boson and other 
SM particles are different from those in the pure SM.  The differences have  
two origins.  First, there is mixing between the $SU(2)_L$ doublet and the 
flavon.  Second, there is sizable mixing between the SM top-quark and the 
heavy fermion $T$. 

Gluon fusion is the most important production channel of the SM Higgs boson 
at the LHC.  In the NP model, the interaction between the SM-like Higgs boson 
and the gluon is mediated by both the SM top-quark and the heavy fermion $T$.  
Denoting the SM and the NP $hgg$ interactions as
\be
c_{hgg}^{SM}hG_{\mu\nu}^aG^{\mu\nu,a},~~~
c_{hgg}^{NP}hG_{\mu\nu}^aG^{\mu\nu,a},
\ee
respectively, we have
\bea
\frac{c_{hgg}^{NP}}{c_{hgg}^{SM}}&=&\frac{\lambda vc_H}{\sqrt2}\left\{\frac{c_Ls_R}{m_t}-\frac{s_Lc_R\tau_T\left[1+\left(1-\tau_T\right)f\left(\tau_T\right)\right]}{m_T\tau_t\left[1+\left(1-\tau_t\right)f\left(\tau_t\right)\right]}\right\}\nonumber\\
&-&\frac{\lambda^\prime vs_H}{\sqrt2}\left\{\frac{s_Ls_R}{m_t}+\frac{c_Lc_R\tau_T\left[1+\left(1-\tau_T\right)f\left(\tau_T\right)\right]}{m_T\tau_t\left[1+\left(1-\tau_t\right)f\left(\tau_t\right)\right]}\right\},
\nonumber\\
\eea
where $s_H(c_H)\equiv\sin\theta_H(\cos\theta_H)$, $\tau_i\equiv4m_i^2/m_h^2$ and 
\be
f\left(\tau\right)=\left\{\begin{array}{l}\arcsin^2\left(\sqrt{1/\tau}\right),~~~~~~~~~~~~~
\tau\geqslant1, \\-\frac{1}{4}
\left[\log\left(\eta_+/\eta_-\right)-i\pi\right]^2,~~~~\tau<1,\end{array}\right.
\ee
where $\eta_\pm\equiv1\pm\sqrt{1-\tau}$.  
The ratio is nearly independent of $\lambda,M$ and $\lambda^\prime$ in the 
3$\sigma$ fit region from EWPO.   In the limit of large fermion mass, we obtain
\bea
\frac{c_{hgg}^{NP}}{c_{hgg}^{SM}}&\to&\frac{\lambda vc_H}{\sqrt2}
\left(\frac{c_Ls_R}{m_t}-\frac{s_Lc_R}{m_T}\right)\nonumber\\
&&-\frac{\lambda^\prime vs_H}{\sqrt2}
\left(\frac{s_Ls_R}{m_t}+\frac{c_Lc_R}{m_T}\right)\nonumber\\
&=& c_H.
\eea
The $h\gamma\gamma$ interaction is also modified in the NP model. 
The contributions
from the light fermions are highly suppressed by the fermion mass, so we 
consider only the contribution from $t$, $T$ and $W^\pm$.  Denoting 
the SM and the new physics (NP) $h\gamma\gamma$ interaction as
\be
c_{h\gamma\gamma}^{SM}hF_{\mu\nu}F^{\mu\nu},~~~
c_{h\gamma\gamma}^{NP}hF_{\mu\nu}F^{\mu\nu},
\ee
and 
\be
\kappa_g=\frac{c_{hgg}^{NP}}{c_{hgg}^{SM}},~~~
\kappa_\gamma=\frac{c_{h\gamma\gamma}^{NP}}{c_{h\gamma\gamma}^{SM}},
\ee
we derive 
\be
\kappa_\gamma=
\frac{N_cQ_f^2A_{1/2}\kappa_g+c_HA_1}
{N_cQ_f^2A_{1/2}+A_1},
\ee
where
\bea
&&A_{1/2}=\tau_t\left[1+\left(1-\tau_t\right)f\left(\tau_t\right)\right],\\
&&A_1=-\frac{1}{2}\left[2+3\tau_W+3\left(2\tau_W-\tau_W^2\right)f\left(\tau_W\right)\right],
\eea
$N_c=3$, and $Q_f=2/3$ are the color factor and charge of the top-quark. 
In the limit of large fermion mass,
\be
\kappa_g=\kappa_\gamma=c_H
\ee
is a very good numerical approximation for this model.

Because $\Phi$ does not couple to the SM fermions except the top-quark, 
all of the other $h\bar ff$ coupling strengths are rescaled by a factor of $c_H$. 
Therefore the NP effects will not change the SM-like Higgs boson
decay branching ratios, but they will change the production cross section. The 
gluon-gluon fusion channel, vector boson fusion (VBF) channel, and the vector boson
associated production (VH) channel are all suppressed by a factor of $c_H^2$.

For the $h\bar tt$ interaction, the ratio between the coupling constant and 
the top-quark Yukawa coupling constant in the SM is
\be
\frac{vs_R}{\sqrt2 m_t}\left(\lambda c_Hc_L-\lambda^\prime s_Hs_L\right).
\ee
It will deviate from $c_H$ by a small amount, but the $\bar tth$ production channel 
has a much smaller cross section than the other three channels.  

The results from a fit of the Higgs boson inclusive cross section $\mu=\sigma/\sigma_{SM}$ 
by the CMS collaboration \cite{CMS-PAS-HIG-13-005} is 
\be
\mu=0.80\pm0.14.
\ee
The result from the ATLAS collaboration 
\be
\mu=1.30\pm0.12({\text{stat}})^{+0.14}_{-0.11}({\text{sys}}).
\ee
would exclude most of the parameter space of the NP model~\cite{ATLAS-CONF-2014-009}.
However, at the 3~$\sigma$ C.L., the region $s_H^2<0.2$ is still allowed.

Although the $h\to VV$ and $h\to f \bar f$ decay branching ratios are not 
changed in this NP model, owing to the universal rescaling factor $c_H$
of $hVV$ and $h\bar ff$ ($h\to t\bar t$ is forbidden because 
$m_h<2m_t$), it is worth checking the $h\to Z^0\gamma$ decay 
branching ratio.  In the NP model, there are additional 
contributions from both the $T$-loop and the $t-T$-loop (FIG. \ref{fig:hza}).
\begin{figure}[!htb]
\includegraphics[scale=0.45,clip]{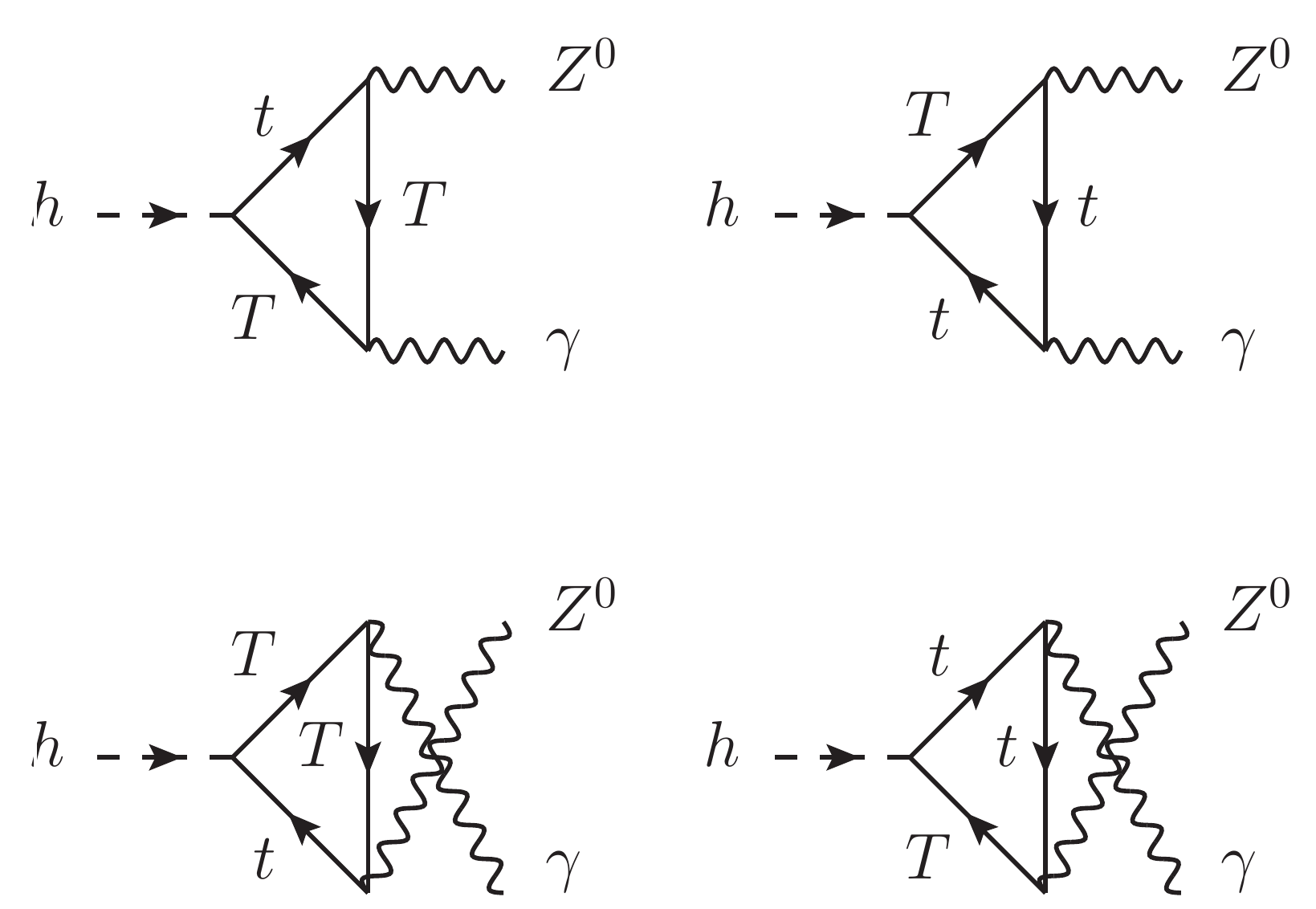}
\caption{Feynman diagrams for the additional contributions to $h\to Z^0\gamma$.  
Both $t$ and $T$ appear in the fermion loops.  
\label{fig:hza} }
\end{figure}
The contributions from the $t$-loop and the $T$-loop can be read out from the rescaling 
of the SM amplitude.  The additional contribution from the Feynman diagrams 
shown in FIG. \ref{fig:hza} must be calculated independently.  The effective operator can 
be written as $c_{hZ\gamma}hZ_{\mu\nu}
F^{\mu\nu}$, where $Z_{\mu\nu}$ and $F_{\mu\nu}$ are the field strengths of the $Z^0$
and the electromagnetic field, respectively. The partial decay width of the Higgs boson
is 
\be
\Gamma=\frac{|c_{hZ\gamma}|^2m_h^3}{8\pi}\left(1-\frac{m_Z^2}{m_h^2}\right)^3 .
\ee
In the SM, there are contributions from the 
fermion loops and the $W^\pm$ loop.  The contribution from the top-quark loop is 
\bea
c_{hZ\gamma}^{SM,t}&=&\frac{\alpha g Q_t\left(8s^2_W-3\right)}
{16\pi m_Ws_Wc_W}\biggl\{\frac{\tau_t\lambda_t}{\tau_t-\lambda_t}\nonumber\\
&&+2m_t^2C_0\left(m_h^2,m_Z^2,0,m_t^2,m_t^2,m_t^2\right)\nonumber\\
&&+\frac{2m_t^2\tau_t\lambda_t}{\tau_t-\lambda_t}
C_0\left(m_h^2,m_Z^2,0,m_t^2,m_t^2,m_t^2\right)\nonumber\\
&&+\frac{\tau_t^2\lambda_t}{\left(\tau_t-\lambda_t\right)^2}\bigl[B_0\left(m_Z^2,m_t^2,m_t^2\right)\nonumber\\
&&-B_0\left(m_h^2,m_t^2,m_t^2\right)\bigr]\biggr\},
\eea
where $\lambda_t\equiv4m_t^2/m_Z^2$, $B_0$ and $C_0$ are the standard
Pasarino-Veltman functions.  We use the $ht\bar t$ coupling constant 
$y_t=\left(gm_t\right)/\left(2m_W\right)$ in the SM.  The contributions from 
the pure top and $T$ loops in the NP case are
\bea
c_{hZ\gamma}^{NP,t}&=&\frac{\sqrt2  m_Ws_R
\left(8s^2_W-3c_L^2\right)}{gm_t\left(8s_W^2-3\right)}\nonumber\\
&&\times\left(\lambda c_Lc_H
-\lambda^\prime s_Ls_H\right) c_{hZ\gamma}^{SM,t},\\
c_{hZ\gamma}^{NP,T}&=&-\frac{\sqrt2 m_Wc_R
\left(8s^2_W-3s_L^2\right)}{gm_T\left(8s_W^2-3\right)}\nonumber\\
&&\times\left(\lambda s_Lc_H
+\lambda^\prime c_Ls_H\right)c_{hZ\gamma}^{SM,t}\left(t\to T\right).
\eea
The new contributions from the $t - T$ mixing loops are
\bea
c_{hZ\gamma}^{NP,tT}&=&-\frac{3\alpha\lambda Q_tc_Hc_Ls_L}{4\sqrt2 \pi
s_Wc_W\left(m_h^2-m_Z^2\right)^2}\biggl\{m_T\left(m_h^2-m_Z^2\right)\nonumber\\
&&\times \left[c_Lc_R\left(m_h^2-2m_T^2-m_Z^2\right)+2m_tm_Ts_Ls_R\right]
\nonumber\\
&&\times C_0\left(m_h^2,m_Z^2,0,m_T^2,m_t^2,m_T^2\right)-m_t\left(m_h^2-m_Z^2\right)\nonumber\\
&&\times\left[s_Ls_R\left(m_h^2-2m_t^2-m_Z^2\right)+2m_tm_Tc_Lc_R\right]\nonumber\\
&&\times C_0\left(m_h^2,m_Z^2,0,m_t^2,m_T^2,m_t^2\right)-2\left(m_h^2-m_Z^2\right)\nonumber\\
&&\times\left(m_Tc_Lc_R-m_ts_Ls_R\right)-2m_Z^2\bigl(m_Tc_Lc_R\nonumber\\
&&-m_ts_Ls_R\bigr)\bigl[B_0\left(m_h^2,m_t^2,m_T^2\right)\nonumber\\
&&-B_0\left(m_Z^2,m_t^2,m_T^2\right)\bigr]\biggr\}\nonumber\\
&&+\frac{3\alpha\lambda^\prime Q_ts_Hc_Ls_L}{4\sqrt2 \pi
s_Wc_W\left(m_h^2-m_Z^2\right)^2}\biggl\{m_T\left(m_h^2-m_Z^2\right)\nonumber\\
&&\times \left[s_Lc_R\left(m_h^2-2m_T^2-m_Z^2\right)-2m_tm_Tc_Ls_R\right]
\nonumber\\
&&\times C_0\left(m_h^2,m_Z^2,0,m_T^2,m_t^2,m_T^2\right)+m_t\left(m_h^2-m_Z^2\right)\nonumber\\
&&\times\left[c_Ls_R\left(m_h^2-2m_t^2-m_Z^2\right)-2m_tm_Ts_Lc_R\right]\nonumber\\
&&\times C_0\left(m_h^2,m_Z^2,0,m_t^2,m_T^2,m_t^2\right)-2\left(m_h^2-m_Z^2\right)\nonumber\\
&&\times\left(m_Ts_Lc_R+m_tc_Ls_R\right)-2m_Z^2\bigl(m_Ts_Lc_R\nonumber\\
&&+m_tc_Ls_R\bigr)\bigl[B_0\left(m_h^2,m_t^2,m_T^2\right)\nonumber\\
&&-B_0\left(m_Z^2,m_t^2,m_T^2\right)\bigr]\biggr\}.
\eea
The analytic formulas of the Passarino-Veltman functions 
can be found in \cite{'tHooft:1978xw,Passarino:1978jh}. The correction from 
NP is comparable to the contribution from the SM fermion loops.
Because the partial width is not rescaled by just $c_H^2$ but 
has a more complicated behavior, the branching ratio is changed 
by the NP. 
We show the ratio between the ${\text{Br}}\left(h\to 
Z^0\gamma\right)$ in the NP and the SM in FIG. \ref{fig:hza2}.
\begin{figure}[!htb]
\includegraphics[scale=0.35,clip]{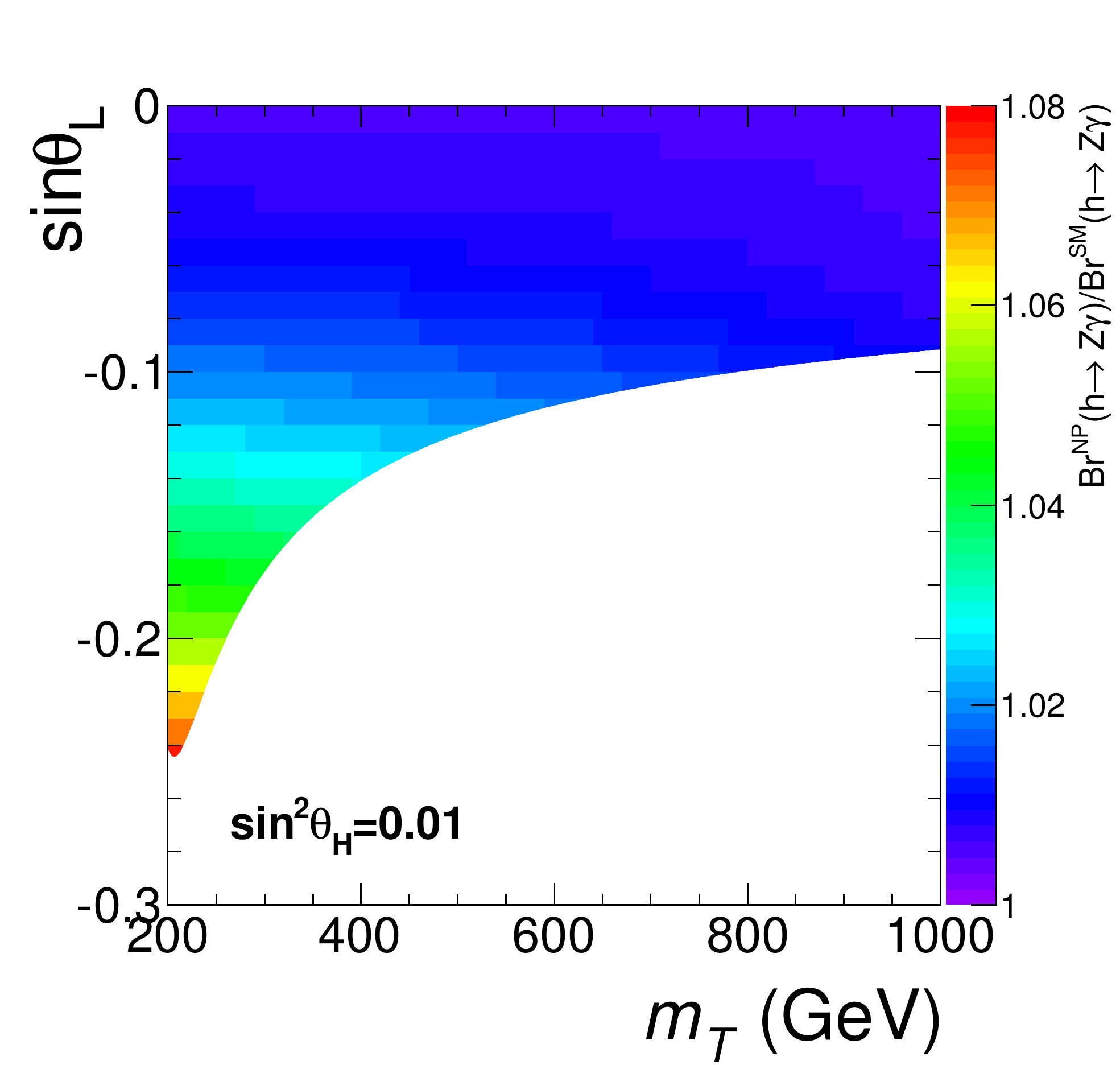}
\includegraphics[scale=0.35,clip]{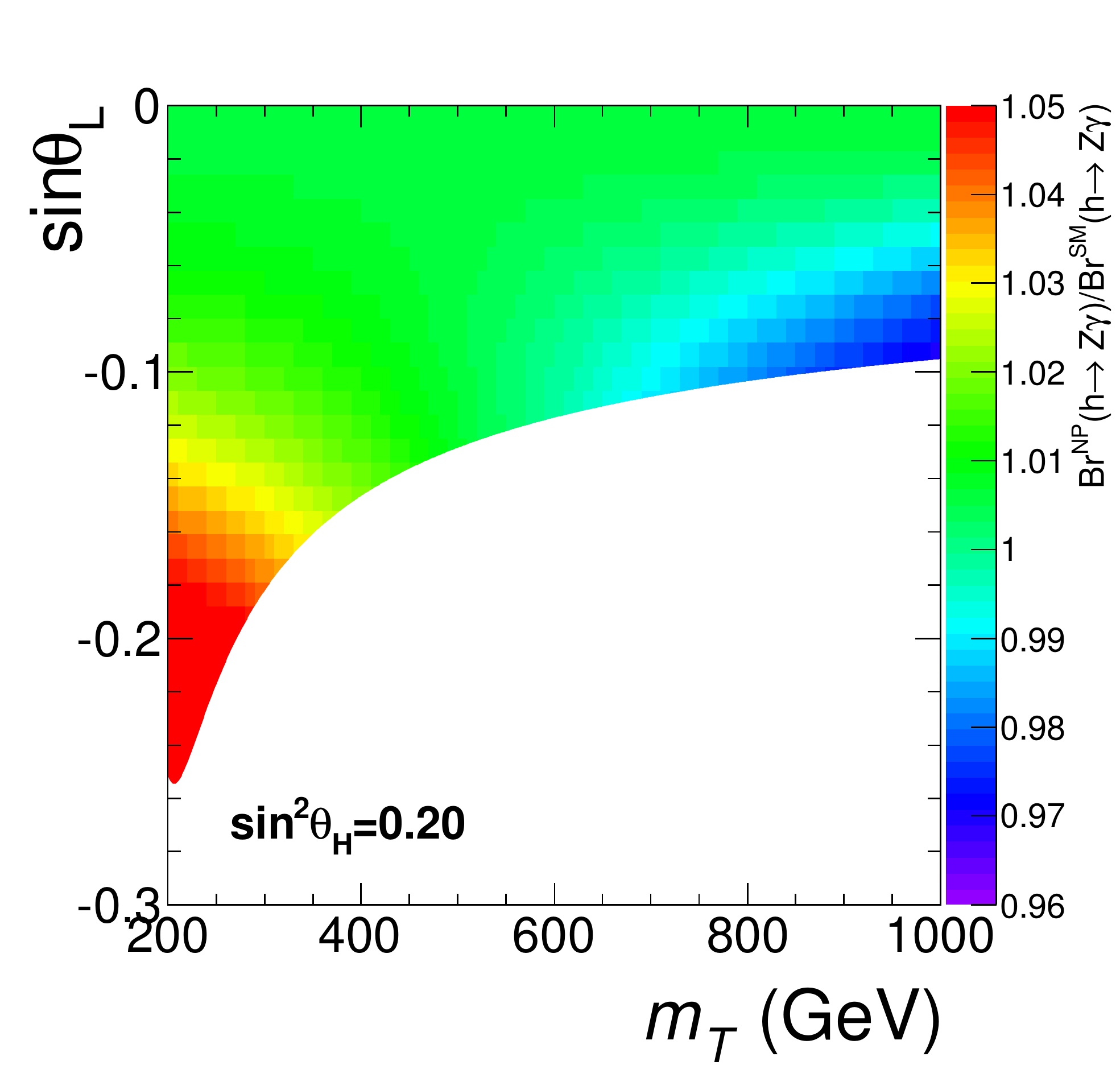}
\caption{The ratio between ${\text{Br}}\left(h\to 
Z^0\gamma\right)$ in the NP model and the SM is shown as 
a function of the mass of the heavy fermion $m_T$ and 
the mixing angle of the left handed fermions $s_L$. We show the 
result in the region where the model can fit the EWPO at 3 $\sigma$ C.L..
We choose 
$m_\varphi=300$ GeV and $\lambda_\Phi=1.0$. In the upper
panel, we choose $s^2_H=0.01$. In the lower panel,
$s^2_H=0.20$. 
\label{fig:hza2} }
\end{figure}
As seen in FIG. \ref{fig:hza2}, the NP contribution increases 
${\text{Br}}\left(h\to Z^0\gamma\right)$.  When the $\sin \theta_L$ and $m_T$ 
parameters satisfy the SM EWPO at 3 $\sigma$ C.L., 
the correction to ${\text{Br}}\left(h\to 
Z^0\gamma\right)$ is small.

Last but not least, an important question is how the SM Higgs pair-production 
cross section \cite{Shao:2013bz} is changed in this model of NP.   There are two sources of change.  
The first is from flavon decay.  The flavon can be produced singly at the LHC.  If 
it decays into the $hh$ final state with a sizable decay branching ratio, the $hh$ cross 
section will be enhanced significantly owing to the resonance.  The second source 
comes from corrections to the $h\bar tt$ and $hhh$
vertices.   We leave flavon production and decay to the next section but discuss the 
modifications of the vertices here.

According to the low-energy theorem \cite{Ellis:1975ap,Shifman:1979eb,Kniehl:1995tn,
Gillioz:2012se}, in this NP model
we have at large $M$
\bea
c_{hgg}&=&c_{hgg}^{SM}c_H,\\
c_{hhgg}&=&\frac{c_{hhgg}^{SM}}{2\left(m_T^2-m_t^2\right)^2}
\biggl\{c_H^2\biggl[2\left(m_T^4+m_t^4\right)\nonumber\\
&&-2\lambda^2v^2
\left(m_T^2+m_t^2\right)+\lambda^4v^4\biggr]\nonumber\\
&&-2c_Hs_H\sqrt{\left(2m_T^2-\lambda^2v^2\right)\left(\lambda^2v^2
-2m_t^2\right)}\nonumber\\
&&\times\left(\lambda^2v^2-m_T^2-m_t^2\right)\frac{\lambda^\prime}
{\lambda}\nonumber\\
&&+s_H^2\left(2m_T^2-\lambda^2v^2\right)\left(\lambda^2v^2
-2m_t^2\right)\frac{\lambda^{\prime2}}
{\lambda^2}
\biggr\},
\eea  
where $c_{hhgg}$ is the coupling constant of the interaction $hhG^{a}_{\mu\nu}
G^{\mu\nu,a}$.   
In the allowed parameter space $\lambda v\to \sqrt2m_t+0^+$,
\be
c_{hhgg}=c_{hhgg}^{SM}c_H^2
\ee 
is therefore a very 
accurate approximation. We 
use this approximation in our numerical calculation.  Moreover, 
\be
\lambda_{hhh}=\frac{m_h^2}{2v}\left(c_H^3-s_H^3
\sqrt{\frac{\lambda_\Phi v^2}{m_\varphi^2c_H^2+m_h^2s_H^2}}\right),
\ee
where $\lambda_{hhh}$ is the trilinear coupling strength of the SM-like Higgs boson.

It was argued recently that a contribution from a higher dimensional 
operator such as $\left(H^\dagger H\right)\bar Q_L\tilde Ht_R$ might be important 
in new physics models \cite{Chen:2014xra}. In the present model, 
such higher dimensional operators can be 
generated when we integrate out the heavy fermion and gauge boson degrees of freedom  
at the cutoff scale $\Lambda\gtrsim10$ TeV (FIG. \ref{fig:effect_tthh}).
\begin{figure}[!htb]
\includegraphics[scale=0.35,clip]{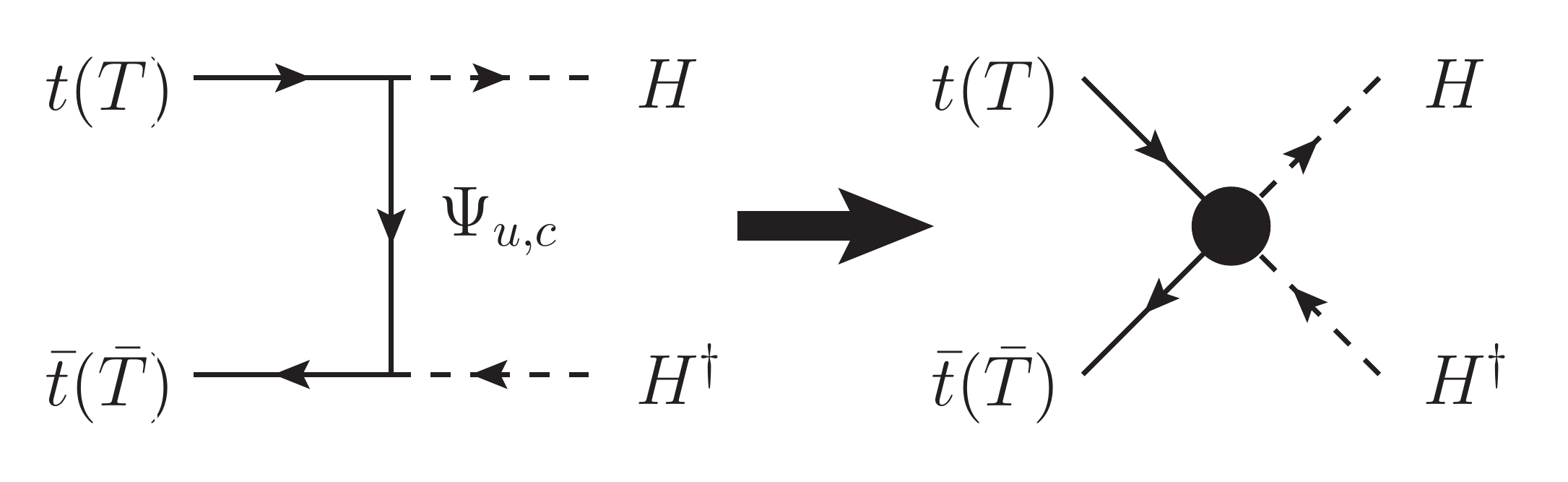}
\caption{The Feynman diagram by which the $\left(H^\dagger H\right)\bar Q_L\tilde Ht_R$
operator can be generated. The $\Psi_{u,c}$ are the heavy fermion eigenstates of the
first and the second generations in the UV completion of the flavor symmetry model 
with inverted hierarchy.}
\label{fig:effect_tthh}
\end{figure}
These operators are suppressed by the mass of $\Psi_{u,c}$, the heavy fermion 
eigenstates of the first and the second generations in the UV completion of the flavor 
symmetry model with inverted hierarchy, and also by a second power of the flavor changing
interaction strength between the third and the first (or second) generation. Since the
mass of $\Psi_{u,c}$ is of the order of $Mv/m_{c,u}\gtrsim100$ TeV, the contribution
from these operators will be small, and we do not include them in our calculation.

We use the program HPAIR \cite{spira:hpair} to calculate the NLO cross section including
the vertex corrections.  In the 
pure SM case, $\sigma_{hh} = 34$~fb at 14 TeV.  The result in the NP case is shown in 
FIG. \ref{fig:hhsm}.  In most of the $\theta_H, m_\varphi$ space, the SM Higgs pair production 
cross section is nearly independent of the mass of the flavon for small 
$\lambda_{\Phi}$.
 Without 
including flavon decays into $hh$, $\sigma_{hh}$ is suppressed for large $s_H^2$.
\begin{figure}[!htb]
\includegraphics[scale=0.36,clip]{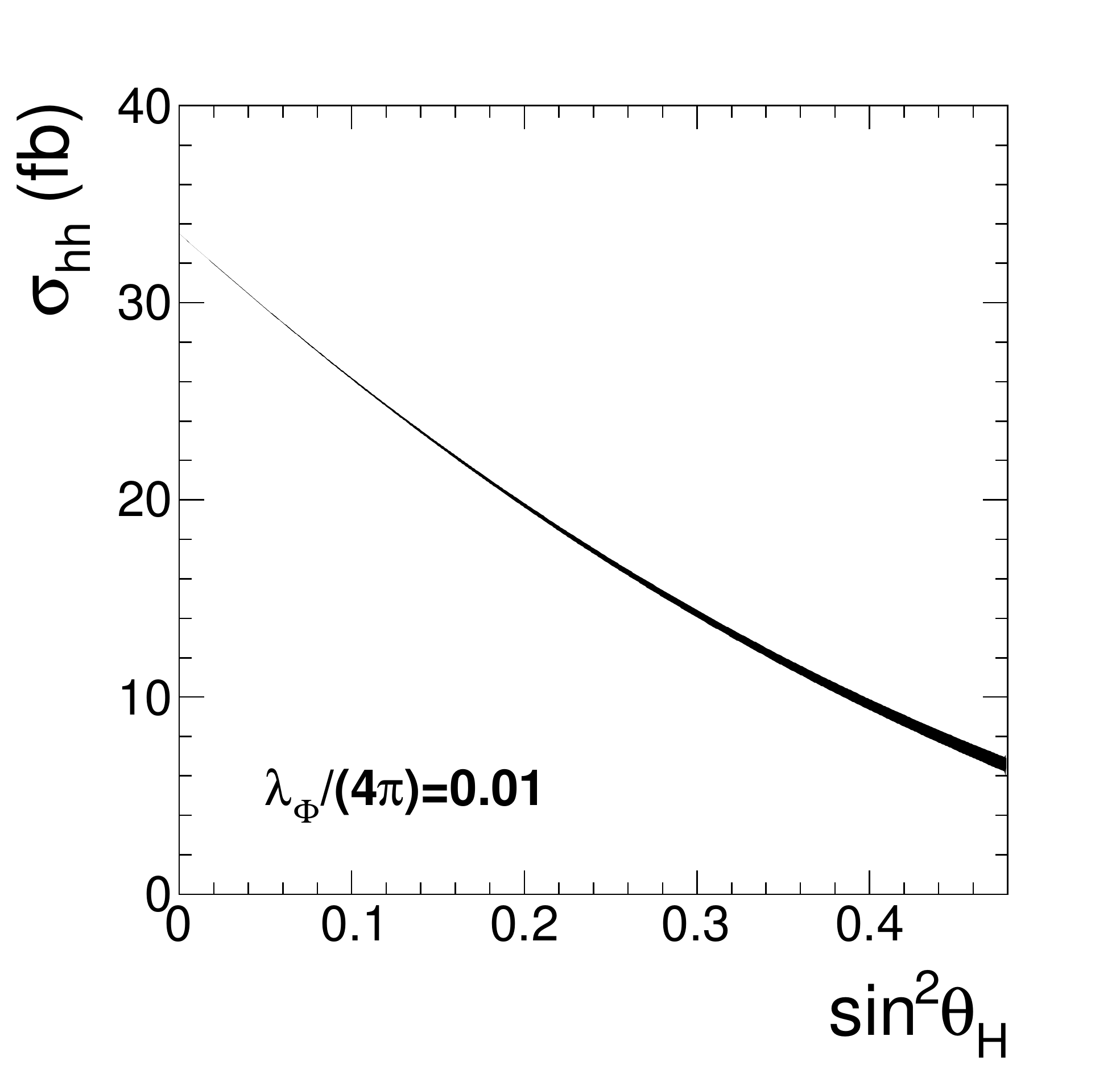}
\caption{The SM-like Higgs pair production cross section in the NP 
model (without the flavon resonance effect). We choose 
$\lambda_\Phi/\left(4\pi\right)=0.01$ in the calculation.
\label{fig:hhsm} }
\end{figure}

To summarize this section, we investigated the production cross section and the decay 
properties of the SM-like Higgs boson $h$ in the gauged flavor symmetry model with inverted 
hierarchy.   We showed that the coupling strengths of
most of the vertices are just rescaled by a factor $c_H$.  We 
checked the loop induced interactions also.  The $hgg$ and $h\gamma
\gamma$ vertices are rescaled by the factor $c_H$ in the heavy fermion limit.
The $hZ^0\gamma$ vertex deviates from the simple $c_H$ rescaling, but the 
deviation is not huge.  The inclusive Higgs production cross section is suppressed 
by a factor $c_H^2$, allowed by the LHC data at 7 and 8 TeV.  The 
decay branching ratios are nearly unchanged relative to the SM since every sizable 
partial width is changed by an overall factor $c_H^2$.   Finally, we computed the $hh$ production 
cross section in this NP model (without the contribution from the flavon resonance decay).

\section{Flavon phenomenology at the LHC}
\label{sec:lhc}
In the previous section, we explored the influence of the NP model on 
SM-like Higgs boson physics at the LHC.  In this section, we investigate searches  
for the flavon $\varphi$.  

The flavon is produced dominantly through gluon fusion at the LHC.  The effective 
interaction between $\varphi$ and gluons is
\be
c_{\varphi gg}\varphi G_{\mu\nu}^aG^{\mu\nu,a},
\ee
where 
\bea
\frac{c_{\varphi gg}}{c_{hgg}^{SM}}&=&\frac{\lambda^\prime vc_H}{\sqrt2}
\left\{\frac{s_Ls_R}{m_t}+\frac{c_Lc_R\tau_T\left[1+\left(1-\tau_T\right)f\left(\tau_T\right)\right]}
{m_T\tau_t\left[1+\left(1-\tau_t\right)f\left(\tau_t\right)\right]}\right\}\nonumber\\
&-&\frac{\lambda vs_H}{\sqrt2}
\left\{\frac{c_Ls_R}{m_t}-\frac{s_Lc_R\tau_T\left[1+\left(1-\tau_T\right)f\left(\tau_T\right)\right]}
{m_T\tau_t\left[1+\left(1-\tau_t\right)f\left(\tau_t\right)\right]}\right\}. \nonumber\\
\label{eq:sgg0}
\eea
Here $c_{hgg}^{SM}$ is the SM $hgg$ effective coupling constant 
for a Higgs boson with the same mass as the flavon. 
When the flavon is heavy, the heavy fermion limit is not a good approximation  (FIG. \ref{fig:sgg_r}).  
In this work, we calculate the flavon production cross section using the full expression Eq. (\ref{eq:sgg0}). 
A global fit of Higgs boson production cross sections at 7 and 8 TeV by the CMS collaboration,
provides $s_H^2\leqslant0.62$ at $3\sigma$ C.L..  According to the global fit by the ATLAS collaboration,
$s_H^2$ should be smaller than $0.21$ at $3\sigma$ C.L..

In contrast to the gluon fusion case, the flavon cross sections in the VBF and vector boson associated 
production channels, where loop effects do not play a role, are just rescaled by 
a factor of $s_H^2$ relative to the Higgs boson cross sections.  
\begin{figure}[!htb]
\includegraphics[scale=0.35,clip]{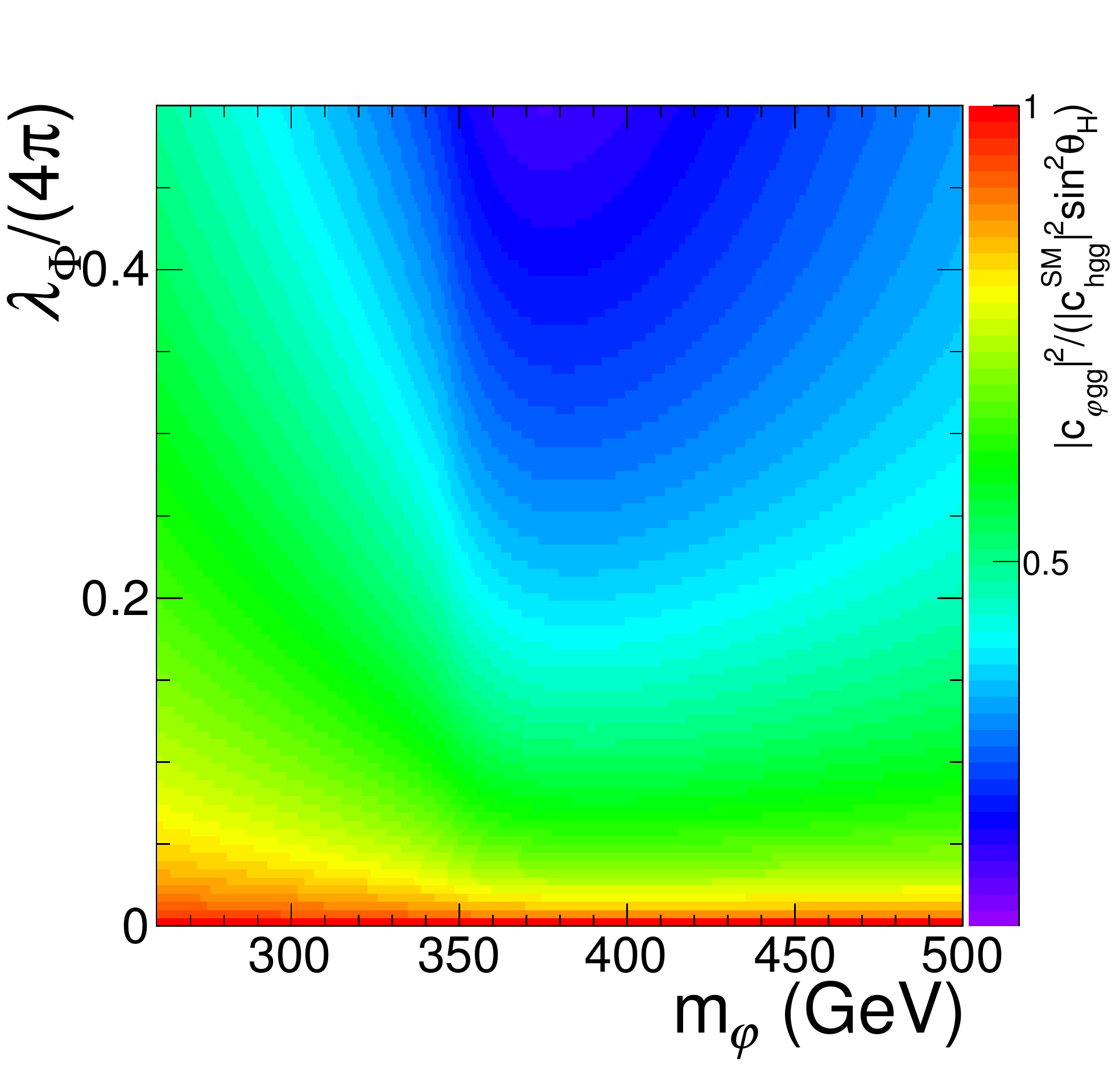}
\caption{The deviation of $|c_{\varphi gg}/c_{hgg}^{SM}|^2$ from the 
heavy fermion limit $s_H^2$ is plotted 
as a function of $\lambda_\Phi$ and the flavon mass $m_\varphi $.}
\label{fig:sgg_r}
\end{figure}
\begin{figure*}[!htb]
\includegraphics[scale=0.28,clip]{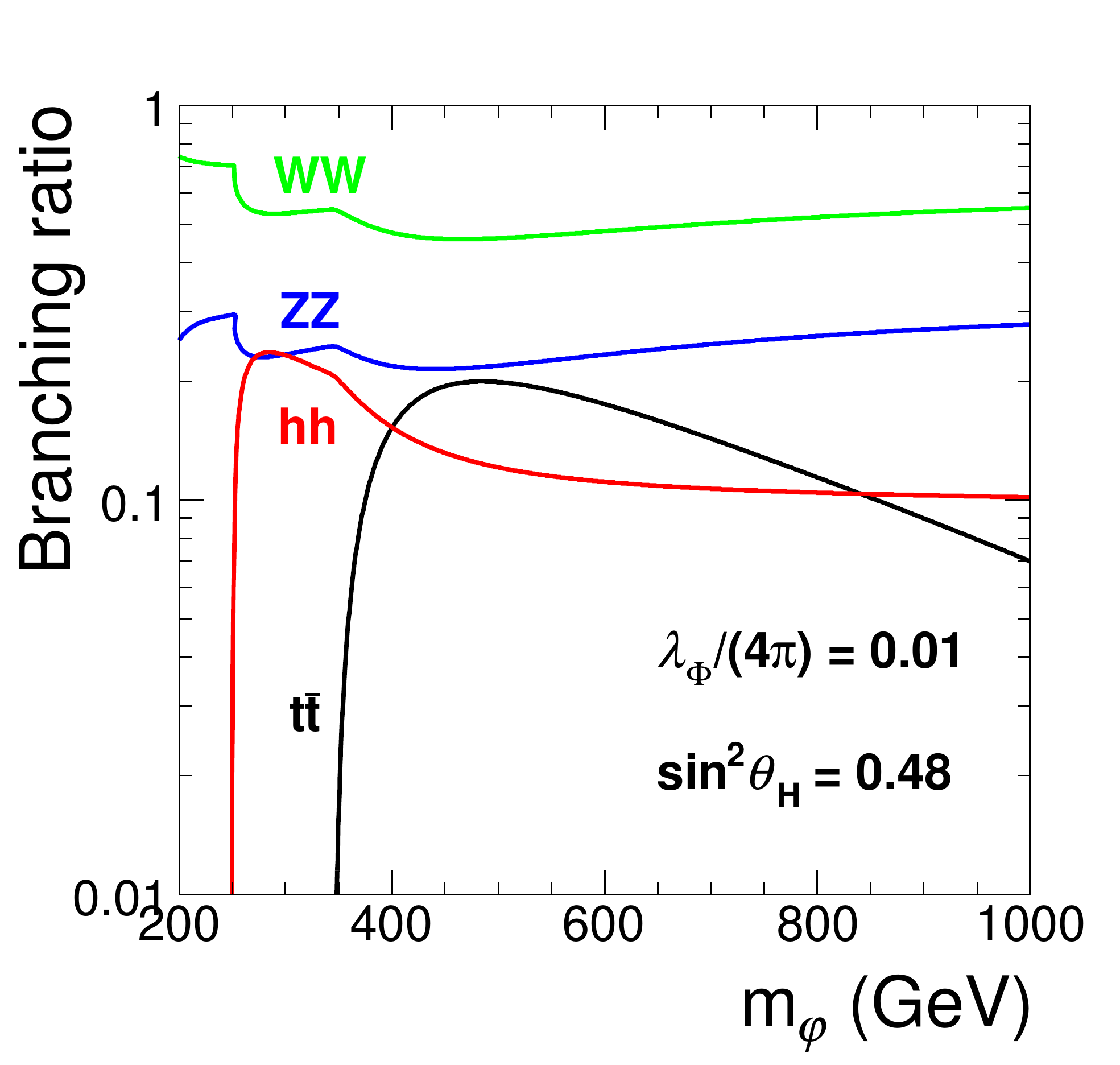}
\includegraphics[scale=0.28,clip]{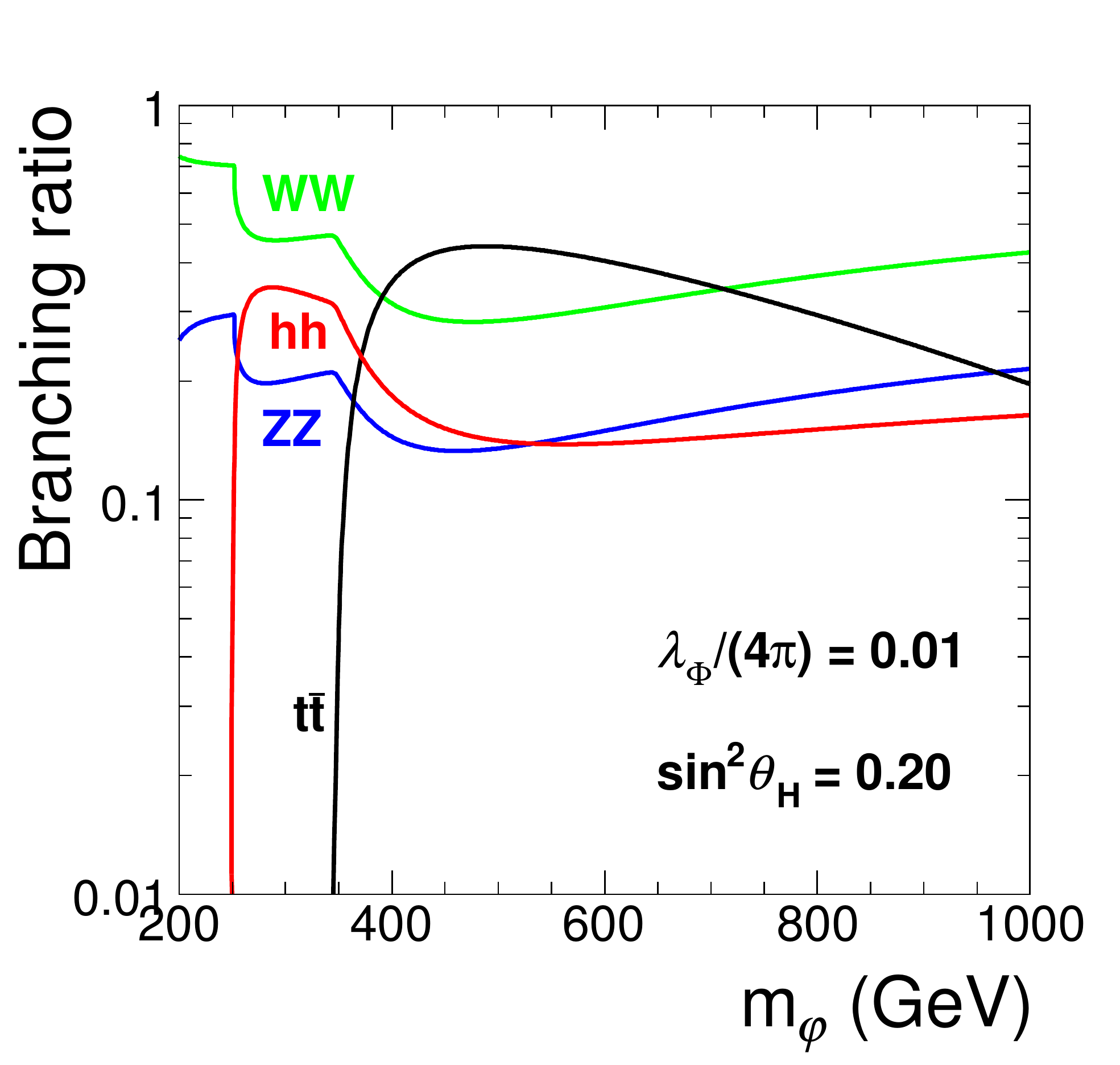}
\includegraphics[scale=0.28,clip]{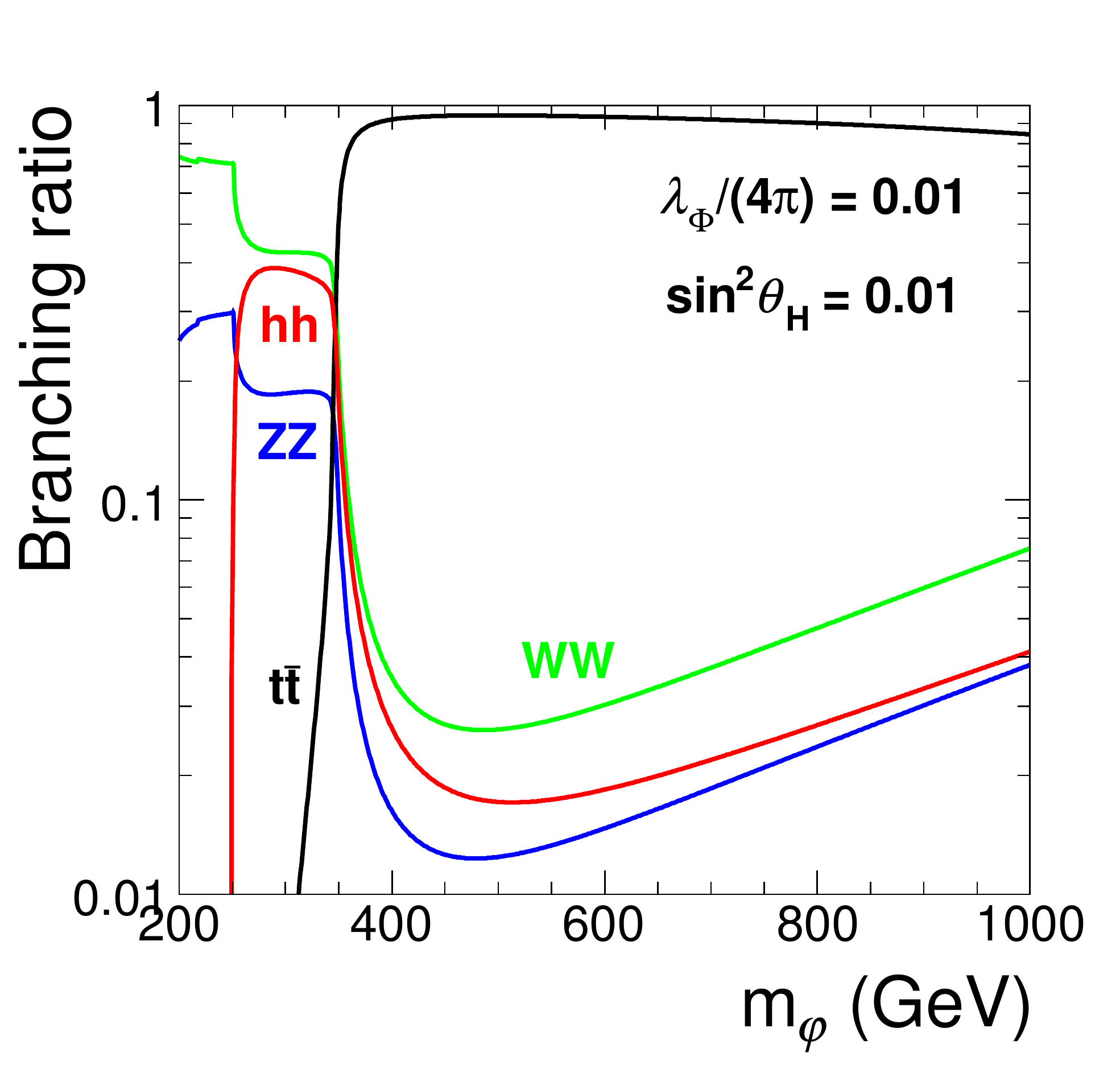}\\
\includegraphics[scale=0.28,clip]{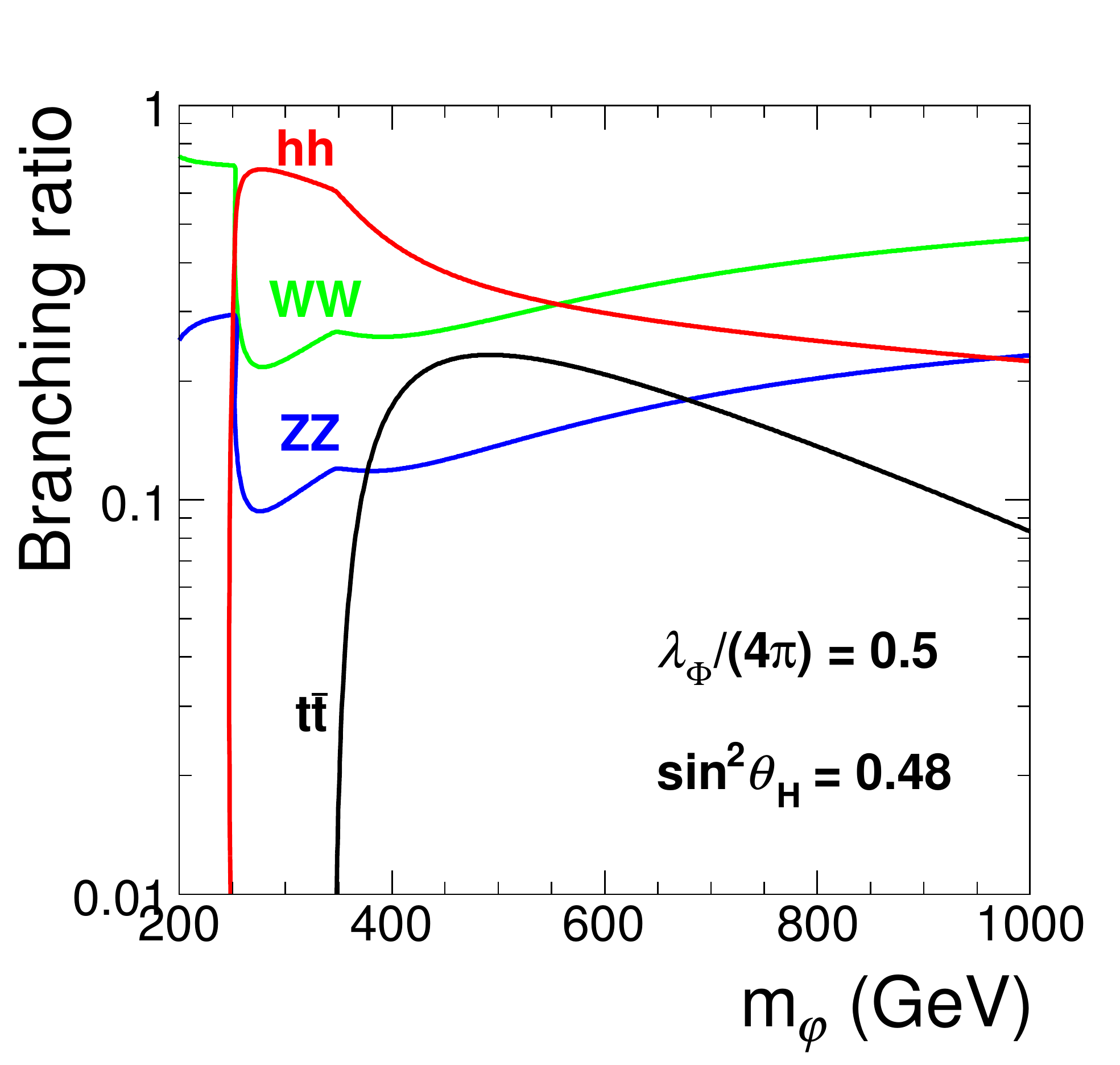}
\includegraphics[scale=0.28,clip]{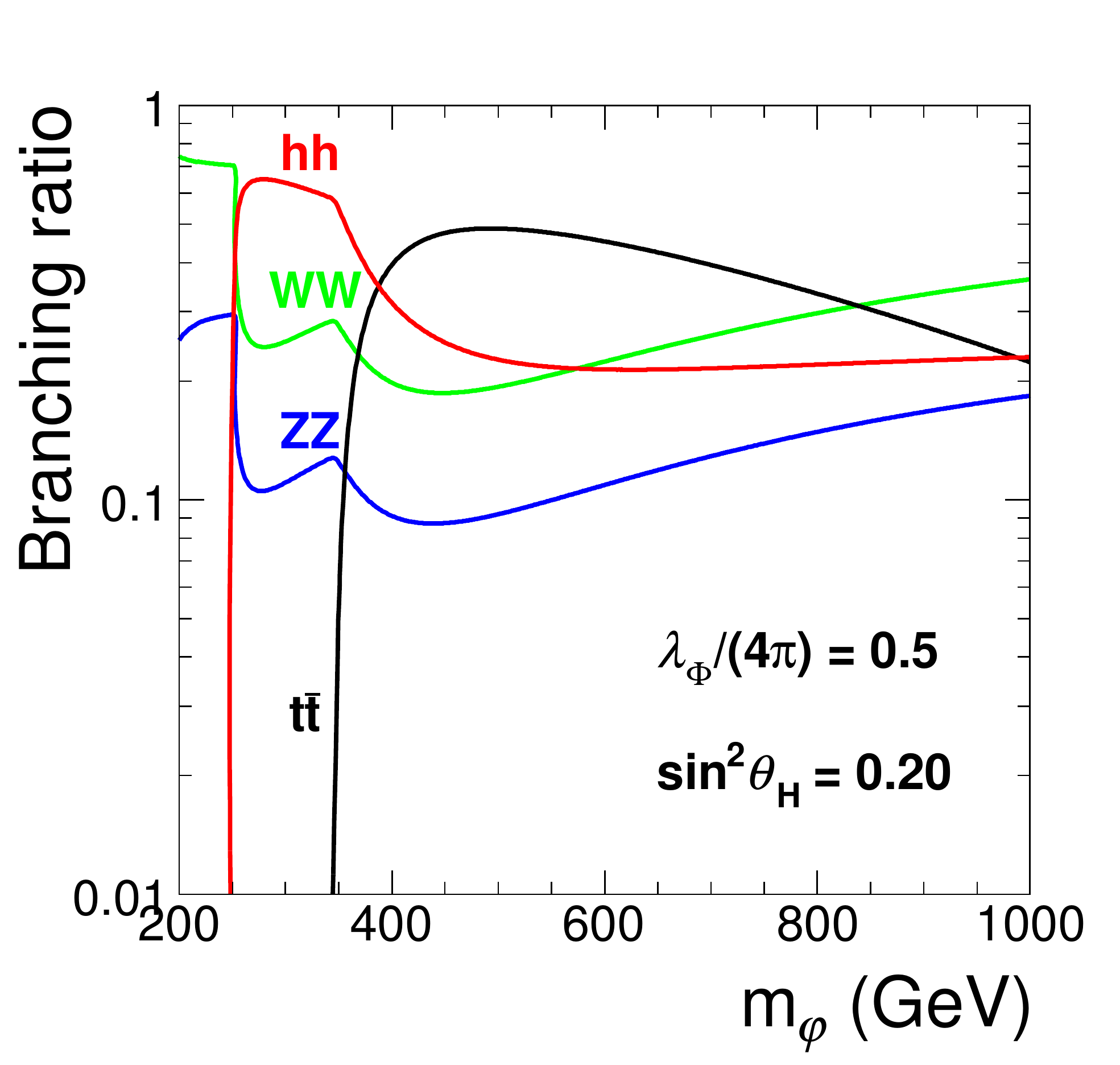}
\includegraphics[scale=0.28,clip]{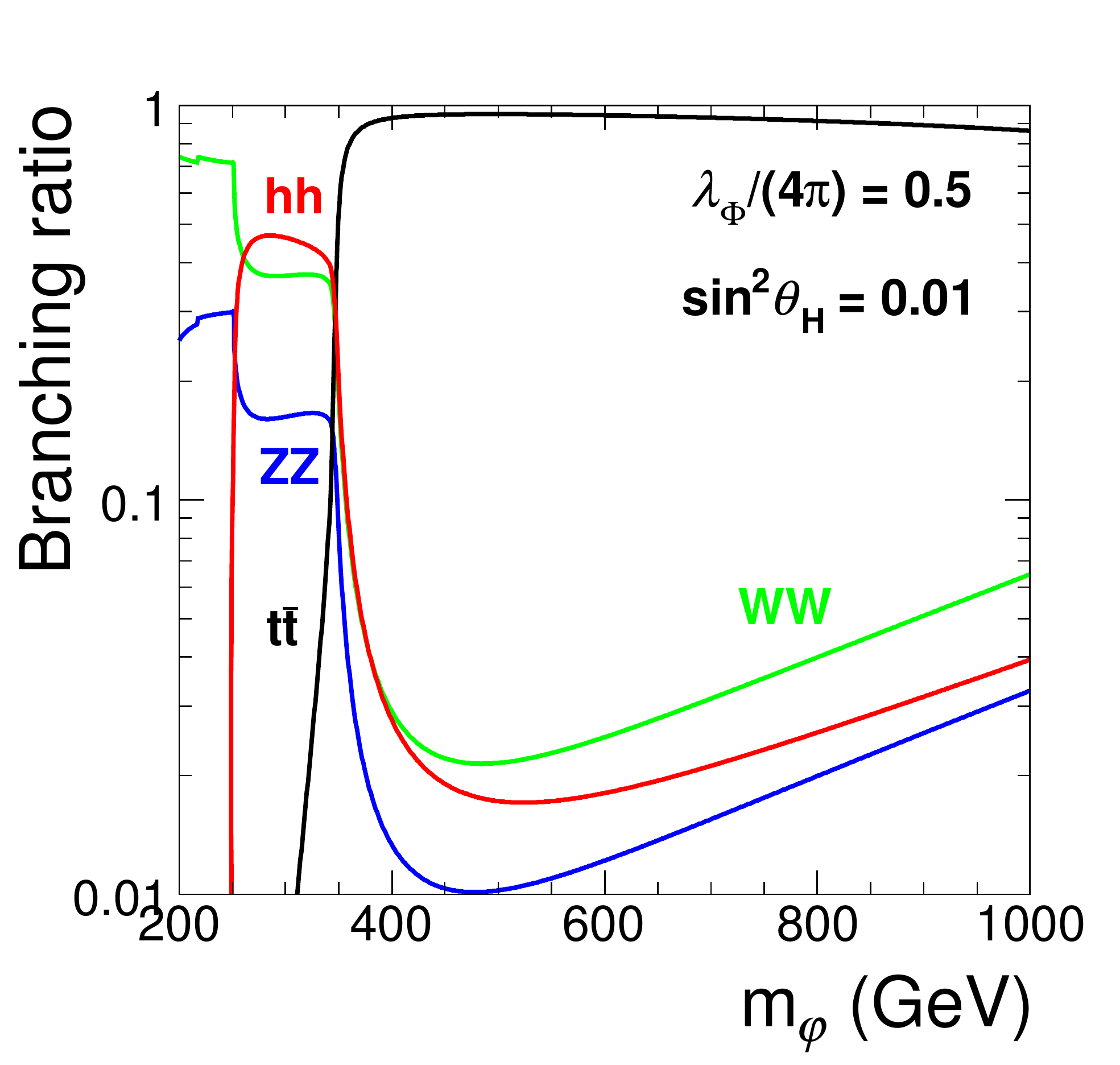}\caption{
The decay branching ratios of the most important $\varphi$ decay channels.
Only the $W^+W^-, Z^0Z^0, hh$ and $t\bar t$ channels are significant in $\varphi$ decay.
In this calculation, we choose $M=2000$ GeV, and $\lambda
=1.1$.
\label{fig:sdecaybr} }
\end{figure*}

The flavon has the same decay modes as the SM Higgs boson.  For a light flavon 
($m_\varphi <2m_t$), the partial decay widths of the regular decay channels 
(except $\varphi  \rightarrow gg, \gamma\gamma, \gamma Z$) are all rescaled by a 
factor of $s_H^2$ which will 
not affect the decay branching ratios of the flavon.  

In addition, it is important to calculate the $\varphi \to hh$ decay width and, for a relatively heavy 
flavon ($2m_t<m_\varphi <m_t+m_T$), the $\varphi \to t\bar t$ decay width. 
For $\varphi \to hh$, we find 
\be
\Gamma\left(\varphi \to hh\right)=\frac{\lambda_{\varphi hh}^2\beta_h}
{8\pi m_\varphi },
\label{eq:shh}
\ee
where $\beta_h\equiv\sqrt{1-4m_h^2/m_\varphi^2}$ and
\bea
\lambda_{\varphi hh}&=&\frac{\left(m_\varphi^2+2m_h^2\right)s_Hc_H}{2v}
\nonumber\\
&&\times\left(c_H+s_H\sqrt{\frac{\lambda_\Phi v^2}{m_\varphi^2
c^2_H+m_h^2s^2_H}}\right).
\eea
In the small mixing limit ($s_H^2\to0$),
we have 
\be
\lambda_{\varphi hh}\to \frac{m_\varphi^2+2m_h^2}{2v}s_H.
\ee
In particular, in the heavy flavon limit ($m_\varphi \gg m_h, m_Z$) with small
$s_H^2$, we have $\Gamma\left(\varphi \to hh\right)=\Gamma\left(\varphi \to Z^0Z^0\right)$
which is a natural expectation of the equivalence theorem \cite{Cornwall:1974km}. 

The contribution from the 3-body decay $\varphi \to h h^*\to h f\bar f$ is \cite{Djouadi:1995gv}
\bea
\Gamma\left(\varphi \to hf\bar f\right)&=&\sum_f\frac{\lambda_{\varphi hh}^2N_cm_f^2}
{256\pi^3 m_\varphi v^2\sqrt{z-1}}\nonumber\\
&\times&\biggl[2\left(5z-4\right)\arctan\left(\frac{\left(z-4\right)\sqrt{z-1}}{4-3z}\right)\nonumber\\
&+&\left(z-4\right)\sqrt{z-1}\left(4+\log\frac{4}{z}\right)\biggr] .
\eea
It is numerically small ($z\equiv4m_h^2/m_\varphi^2$). 

We may calculate the partial width for $s\to t^{(*)}\bar t^{(*)}$ (on-shell 
or off-shell top quarks)
by rescaling the SM 
$h\to t^{(*)}\bar t^{(*)}$ with the factor
\be
\frac{s_R^2v^2}{2m_t^2}\left(\lambda c_Hc_L-\lambda^\prime s_Hs_L\right)^2.
\ee
Since there is no $s_H^2$ suppression in the $\varphi t\bar t$ vertex, the 
flavon will decay into $t\bar t$ with a large branching ratio at small $s_H$
if allowed by phase space.
However, the flavon production cross section is highly suppressed 
by $s_H^2$ in this region.   The signal will be hidden under the SM $t\bar t$ 
background making this signal for the flavon hard to find.

For a heavy flavon, as noted before, the heavy fermion limit is not a good approximation.  
We calculate the branching 
ratios of the loop-induced processes ($\gamma\gamma, gg, Z^0\gamma$) using the 
exact formula.  The calculation is straightforward,  but the result depends on all 
parameters.  For $m_\varphi >160$ GeV, the contribution from the loop induced channels 
is negligibly small, and the most important decay modes are $b\bar b,
t\bar t, W^+W^-, Z^0Z^0$, and $hh$.  These are the decay modes we must 
consider since we treat only $m_\varphi <m_T$ and $m_\varphi<2m_{Z_T}$ in this work.    
The decay branching ratios for the dominant decay channels
can be found in FIG. \ref{fig:sdecaybr}.  
For small $s_H$ ($s_H^2 = 0.01$), $\varphi \rightarrow t \bar{t}$ dominates $\varphi$ decay 
when the channel is open.  This result arises because the $\varphi t \bar{t}$ interaction comes from 
the flavon-top interaction in Eq. \ref{eq:ope}.

In summary, $\varphi \to hh$ is an important decay channel of the flavon, and it might be used to 
discover the flavon at the LHC.

Flavon searches at the LHC can focus on the SM Higgs-like decay 
channels ($Z^0Z^0$, $W^+W^-$) and on the light Higgs boson pair 
decay channel.  Although the SM Higgs-like decay channels of the flavon are 
suppressed by the presence of the $hh$ decay channel, one should  
nevertheless check them carefully.  In the remainder of this section, we 
examine constraints on the flavon from data at 7 and 8 TeV at the LHC 
and study flavon phenomenology at 14 TeV.

\subsection{Limits from the 7 and 8 TeV LHC data}

At 7 and 8 TeV, the strongest limit on the flavon is provided by the 
$Z^0Z^0\to 2\ell2\ell'$ channel in heavy SM Higgs boson 
searches \cite{ATLAS-CONF-2013-013,Chatrchyan:2013mxa}.  The CMS 
collaboration also investigated the $hh$ channel \cite{CMS-PAS-HIG-13-025,
CMS-PAS-HIG-13-032}.
For a heavy enough flavon, $t\bar t$ resonance searches may also constrain the 
parameters \cite{ATLAS-CONF-2013-052,Chatrchyan:2013lca}.  However, the 
small production cross section makes this constraint weak.  We do not discuss 
it here.

\subsubsection{The $\varphi  \rightarrow Z^0Z^0$ channel}

\begin{figure}[!htb]
\includegraphics[scale=0.35,clip]{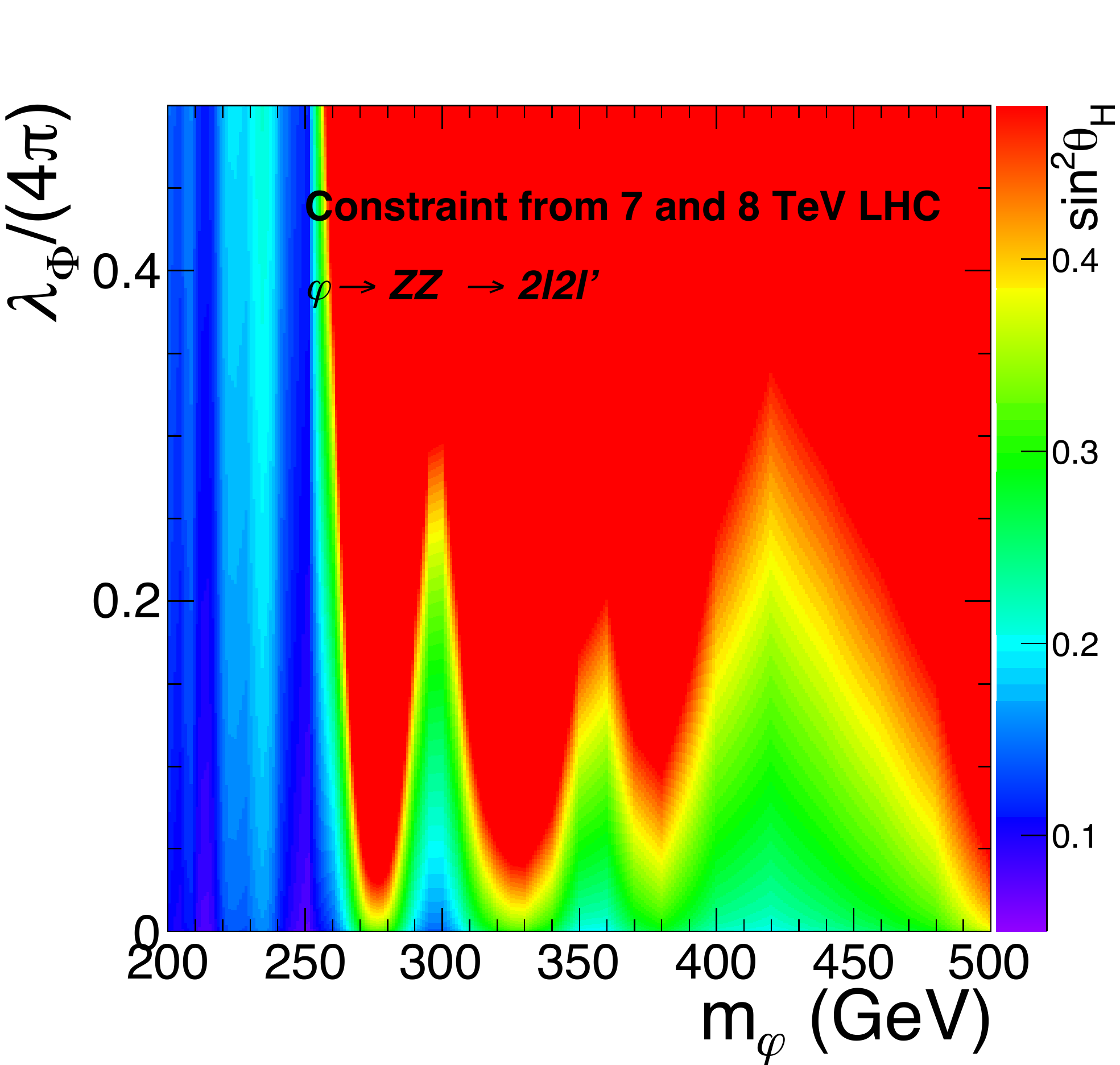}
\caption{The lower bound of the 2$\sigma$ exclusion region of $s_H^2$
from the $Z^0Z^0$ channel heavy SM Higgs boson searches 
at 7 and 8 TeV is shown as a function of 
$\lambda_\Phi$ and $m_\varphi $.  We choose
$M=2000$ GeV in this calculation. In the red region,
the constraint from fitting the Higgs boson inclusive cross section
is stronger. The peaks and valleys in the figure reflect the structure 
of the experimental exclusion bound.
\label{fig:78lhczz} }
\end{figure}

The $Z^0Z^0$ channel has been examined by the ATLAS and  
CMS collaborations.  ATLAS uses data corresponding 
to integrated luminosities of 4.6 fb$^{-1}$ and 20.7 fb$^{-1}$ at 7
TeV and 8 TeV, respectively \cite{ATLAS-CONF-2013-013} (the 
constraint on a heavy SM Higgs boson is from the 8 TeV data only). 
The CMS collaboration uses data corresponding 
to integrated luminosities of 5.1 fb$^{-1}$ and 19.7 fb$^{-1}$ at 7
TeV and 8 TeV, respectively \cite{Chatrchyan:2013mxa}.  We
use these results to place constraints on the flavon.  

For the ATLAS analysis, we use the gluon fusion channel because this channel 
is the dominant production channel for $\varphi$ at the LHC.  We use the expressions 
presented in this paper to rescale the Higgs boson production cross sections 
in \cite{Dittmaier:2011ti} to get a NNLO QCD result.  The numerical results are  
shown in FIG. \ref{fig:78lhczz} where we present the lower bound of the 
2$\sigma$ exclusion region of $s_H^2$ from heavy SM Higgs boson searches at 7 
and 8 TeV as a function of the flavon self-iteraction parameter $\lambda_\Phi$ and its 
mass $m_\varphi$.   In the blue
region, values of $s_H^2$ above $0.2$ are 
excluded at $2\sigma$. 

\subsubsection{The $\varphi  \rightarrow hh$ channel}
The CMS collaboration searched for a heavy scalar (pseudo-scalar) decaying 
into $hh$ ($Z^0h$) at 8 TeV with 19.5 fb$^{-1}$ of data \cite{CMS-PAS-HIG-13-025,
CMS-PAS-HIG-13-032}. 
Multilepton events with or without a diphoton and  $b\bar b\gamma\gamma$
in the final state were 
used in those searches. This channel could be important for a flavon search when 
the $\varphi \to hh$ channel opens.  However, with current luminosity, it does not yield a 
stronger constraint on NP than the SM Higgs global-fit. The ATLAS collaboration 
searched for a TeV-scale resonance decaying into $hh$ at 8 TeV with 19.5 fb$^{-1}$ 
of data using the $b\bar bb\bar b$ final state \cite{ATLAS-CONF-2014-005} and 
$b\bar b\gamma\gamma$ final state \cite{Aad:2014yja}.

It might seem strange that the $hh$ channel cannot provide a stronger constraint 
even when the $Z^0Z^0$ channel is highly suppressed, but it is understandable 
once we check the result carefully.  The 2$\sigma$ level lower bound on $s_H^2$ 
from the $Z^0Z^0$ channel is $0.1-0.2$.  However, the upper bound on 
$\sigma\left(pp\to \varphi +X\to hh+X\right)$ is several picobarns, which is already 
above the SM heavy Higgs boson cross section in that mass region.  Thus it cannot 
provide a constraint on $s_H^2$. 

\subsection{Searches for $\varphi$ at 14 TeV}
We investigate the possibility of discovering a flavon at 14 TeV with 100 fb$^{-1}$ integrated 
luminosity.

\subsubsection{$Z^0Z^0$ channel}
There are simulations by the ATLAS collaboration
\cite{ATLAS-PHYS-PUB-2013-016} and the CMS collaboration
\cite{CMS-PAS-FTR-13-024} for this channel. 
\begin{figure}[!htb]
\includegraphics[scale=0.35,clip]{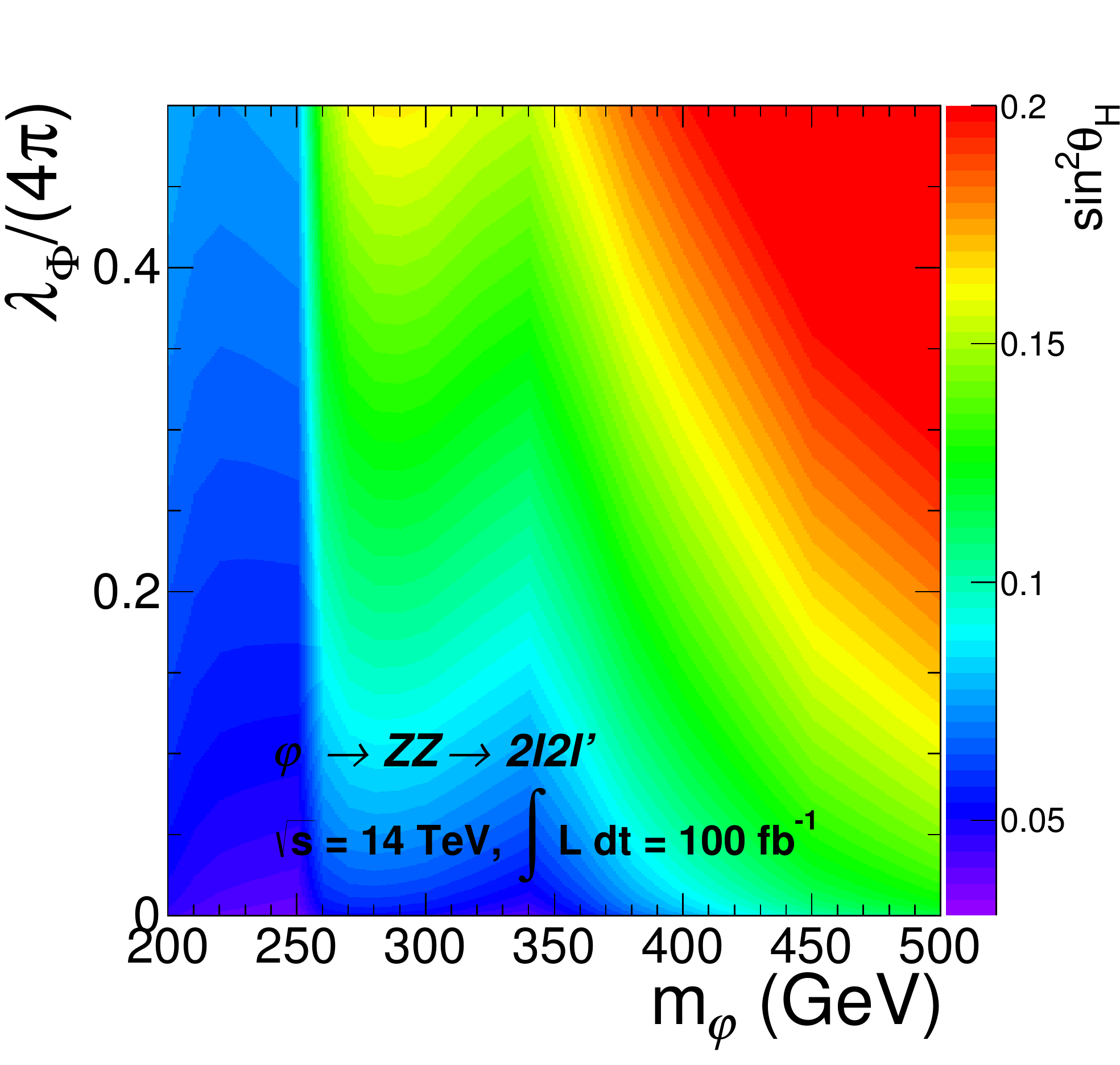}
\caption{The expected lower bound of the 2$\sigma$ exclusion region of $s_H^2$
from searches for heavy scalar decays to $Z^0Z^0$
at the LHC at 14 TeV 
with 100 fb$^{-1}$ integrated luminosity is shown as a function of 
$\lambda_\Phi$ and $m_\varphi $.  We choose
$M=2000$ GeV in this calculation.
\label{fig:14lhczz} }
\end{figure}
We rescale their upper bounds to 100 fb$^{-1}$ integrated luminosity 
by $\sqrt{\mathcal{L}_{\text{int}}/100{\text{fb}^{-1}}}$.
\begin{figure*}[!htb]
\includegraphics[scale=0.90,clip]{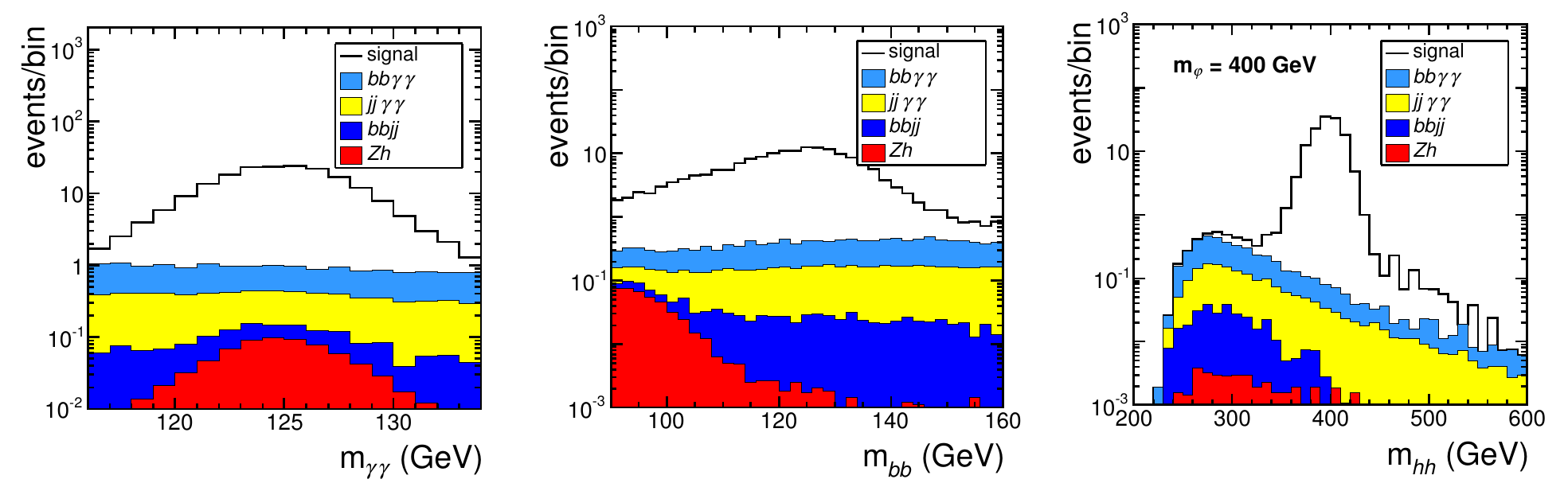}
\caption{The reconstructed diphoton, $b\bar b$, and $hh$ mass distributions 
are shown at 14 TeV with 100 fb$^{-1}$ integrated luminosity.
We choose $m_\varphi =$400 GeV. The total cross section is rescaled to 
the value of the SM like Higgs boson with the same mass. The decay branching ratio
${\text{Br}}\left(\varphi \to hh\right)$ is set to be 100\%. In the left and middle panels, the events
passed the all of the cuts except the $\gamma\gamma$ and $bb$ invariant mass cuts. In the
right panel, the events pass all of the cuts. Note that the horizontal scale differs in the three distributions.
\label{fig:s300} }
\end{figure*}
The constraint from this channel is shown in FIG \ref{fig:14lhczz}. 
When $m_\varphi  <2m_h$, it is evident from the left blue and green band in the figure 
that the $Z^0Z^0$ channel can provide a very strong constraint on the NP model 
(e.g., all $s_H^2$ greater than $\sim 0.08$ is excluded).  
When $m_\varphi >2m_h$, the constraint on  
$s_H^2$ is at $\mathcal{O}\left(10^{-1}\right)$. In this region of $s_H^2$,
the $hh$ channel 
will be the dominant decay channel of $\varphi$.

\subsubsection{$hh$ channel}
We focus on the $b\bar b\gamma\gamma$ channel and present the results of our 
detailed simulation of the signal and backgrounds.  Efforts have been made to use 
this channel for new resonance searches, for example, 
see \cite{Craig:2013hca,Liu:2013woa,Chen:2013emb,Martin:2014zz}. 
The $b\bar b\tau^+\tau^-$ channel \cite{No:2013wsa} is also useful but we do not discuss it. 
The $b\bar b W^+W^-$ channel, which is useful when the final state SM Higgs 
bosons are boosted \cite{Papaefstathiou:2012qe}, is less 
important because the Higgs bosons are not highly boosted in our case.
There are several irreducible SM backgrounds
\bea
pp&\to& b\bar b \gamma\gamma,\nonumber\\
pp&\to& Z^0h \to b\bar b \gamma\gamma,\nonumber\\
pp&\to& Z^0\gamma\gamma\to b\bar b \gamma\gamma,
\eea
and reducible SM backgrounds
\bea
pp&\to& b\bar b jj~\left(j\to \gamma\right),\nonumber\\
pp&\to& jj\gamma\gamma~\left(j\to b\right),\nonumber\\
pp&\to& t\bar t\to bjj\bar bjj~\left(j\to \gamma\right),\nonumber\\
pp&\to& t\bar t h\to b\ell^+ \nu\bar b\ell^- \bar\nu\gamma\gamma
~\left(\ell^\pm~{\text{missed}}\right).
\eea
We generate the signal and background events at the parton level using 
MadGraph 5 \cite{Alwall:2011uj,Alwall:2014hca} with CTEQ6L1
parton distribution functions (PDF) \cite{Pumplin:2002vw}.  
For signal events, we generate $pp\to \varphi +{\text{n}}j$ 
to n=1. 
All of the parton level signal and the background events
are showered using Pythia6.4 \cite{Sjostrand:2006za}. The MLM matching
scheme \cite{Mangano:2006rw} 
is used to avoid double counting.  Detector effects
are mimicked with PGS4 \cite{pgs}.  Jets are defined in the events with the 
anti-$k_T$ algorithm, with $R=0.4$.  The cross section of the $Z^0h$
process is reweighted to the value suggested by the LHC Higgs Cross
Section Working Group in which the NNLO QCD and NLO EW corrections
have been included \cite{Dittmaier:2011ti}. The $Z^0\gamma\gamma$ and 
$b\bar b \gamma\gamma$ cross sections are reweighted to the NLO QCD
results by mutiplication with a simple $K$-factor \cite{Bozzi:2011en,
Gehrmann:2013sla,Bern:2013bha}. The jet fake rate is rescaled 
by \cite{ATL-PHYS-PUB-2013-009}
\be
\epsilon\left(p_T\right)_{j\to\gamma}=9.3\times10^{-3}\times\exp\left(-\frac{p_T}{27.5~
{\text{GeV}}}\right).
\ee
We require the events to have at least 
two hard isolated photons in the central region, which means 
\be
p_T^\gamma>20 {\text{GeV}},~~\left|\eta^\gamma\right|<2.0,
\ee
and no hard jet or charged lepton in the $\Delta R=0.4$ region around the
photon. The events should also have at least two hard $b$-tagged jets with
\be
p_T^j>40~ {\text{GeV}},~~\left|\eta^j\right|<2.0.
\ee
The average 
$b$-tagging
efficiency is reweighted to 70\% \cite{ATL-PHYS-PUB-2013-009} 
in this analysis. The light flavor ($g, u, d, s$) and charm quark mis-tag rates 
are chosen to be 3.9\% and 25.7\%, respectively.
To suppress the SM $t\bar th$ background, we reject events which 
contain hard isolated charged leptons with
\be
p_T^{\ell}>20~ {\text{GeV}},~~\left|\eta^\ell\right|<2.5, ~~\left(I_{iso}<0.1~{\text
{for}}~\mu^\pm\right)
\ee
and events which have a large missing transverse energy 
\be
\met>30~ {\text{GeV}}.
\ee

We require signal events to satisfy hard cuts designed for the Higgs boson pair 
signal. The leading and subleading photon in the events should satisfy 
\be
\left|m_{\gamma\gamma}-125.4 ~{\text{GeV}}\right|<\Delta m_{h,{\text{cut}}}^{\gamma\gamma}.
\ee
The transverse momentum of the leading and of the sub-leading photon should satisfy
\be
p_T^{\gamma_1}>p_{T,{\text{cut}}}^{\gamma_1},~~p_T^{\gamma_2}>p_{T,{\text{cut}}}^{\gamma_2}.
\ee
We require
the leading and subleading $b$-tagged jets to satisfy
\be
\left|m_{bb}-125.4 ~{\text{GeV}}\right|<\Delta m_{h,{\text{cut}}}^{bb}.
\ee
$\delta\phi_{\gamma b}$, the smallest of $\Delta\phi_{\gamma_1b_1},\Delta\phi_{\gamma_1b_2},\Delta\phi_{\gamma_2b_1},
\Delta\phi_{\gamma_2b_2}$ (the differences between the azimuthal angles of the objects) 
should be less than $\Delta\phi_{\gamma b}$.

The energy resolutions of $b$-jets and photons are obtained from the
$Z^0b\to\mu^+\mu^-b$ and the $Z^0\gamma
\to\mu^+\mu^-\gamma$ processes.  After including the energy 
resolution, we can reconstruct the invariant mass peak of the flavon. 

The values of the $p_{T,{\text{cut}}}^{\gamma_1},p_{T,{\text{cut}}}^{\gamma_1},\Delta m_{h,{\text{cut}}}^{bb},
\Delta m_{h,{\text{cut}}}^{\gamma\gamma}$, $\Delta\phi_{\gamma b}$ and the invariant mass window
of the $hh$ system are chosen with the mass 
of the flavon to get a maximal signal significance (TABLE \ref{tab:cuts_ex} and \ref{tab:cuts_dis}). When the mass of the flavon 
increases, a larger leading photon $p_T$ cut will give larger significance because the final state Higgs 
boson is more boosted. Such a large leading photon $p_T$ cut will suppress the SM background 
so that we can release the diphoton invariant mass cut to include more signal events. 

For a flavon whose mass is relatively small, the $b$-jets (diphoton) from the 
Higgs boson decay are not too collinear. In such a case, a $\Delta\phi_{\gamma b}$ cut can 
suppress the QCD background from $q\bar q\to b\bar b\gamma\gamma$ where the $b$-jets
are from a virtual gluon splitting and the photons are from the initial state radiation (ISR). This 
is the reason a small $\Delta\phi_{\gamma b}$ can give larger significance for light flavons.
For the $m_\varphi\sim 280-290$ GeV region, the production cross section drops and the $\Delta\phi
_{\gamma b}$ cut will remove many signal events and not be helpful for increasing significance 
if we cannot increase the $p_T$ cuts.  Thanks to the ``bump'' structure in the $gg\to \varphi$
cross section around $m_\varphi \sim 2m_t$, there is a larger cross section for that region of flavon mass.
More signal events allow us to use harder $p_T$ cuts and the $\Delta\phi_{\gamma b}$
cut.  For a heavy flavon, the two $b$-jets (diphoton) are more and more collinear and $\Delta\phi_{
\gamma b}$ is no longer a good cut.
\begin{table*}[htdp]
\caption{The combination of cuts which gives the largest significance for excluding the flavon with different mass.}
\begin{center}
\begin{tabular}{|c|c|c|c|c|c|c|}
\hline  Exclusion &$p_{T,{\text{cut}}}^{\gamma_1}$&$p_{T,{\text{cut}}}^{\gamma_2}$&$\Delta m_{h,{\text{cut}}}^{\gamma\gamma}$&$\Delta m_{h,{\text{cut}}}^{bb}$&$\Delta\phi_{\gamma b}$ & $m_{hh}$ window \\\hline
\hline ~~~~$m_\varphi =$ 260 GeV~~~~ & ~~~~30 GeV~~~~ &~~~~20 GeV~~~~&~~~~3.5 GeV~~~~& ~~~~20 GeV~~~~& ~~~~$0.5\pi$~~~~& ~~~~230 GeV $\sim$ 280 GeV~~~~ \\
\hline ~~~~$m_\varphi =$ 270 GeV~~~~ & ~~~~35 GeV~~~~ &~~~~20 GeV~~~~&~~~~3.5 GeV~~~~& ~~~~20 GeV~~~~& ~~~~$0.5\pi$~~~~& ~~~~245 GeV $\sim$ 295 GeV~~~~ \\
\hline ~~~~$m_\varphi =$ 280 GeV~~~~ & ~~~~35 GeV~~~~ &~~~~20 GeV~~~~&~~~~3.5 GeV~~~~& ~~~~15 GeV~~~~& ~~~~$\pi$   ~~~~& ~~~~265 GeV $\sim$ 300 GeV~~~~  \\
\hline ~~~~$m_\varphi =$ 290 GeV~~~~ & ~~~~30 GeV~~~~ &~~~~20 GeV~~~~&~~~~3.5 GeV~~~~& ~~~~15 GeV~~~~& ~~~~$\pi$   ~~~~& ~~~~275 GeV $\sim$ 310 GeV~~~~  \\
\hline ~~~~$m_\varphi =$ 300 GeV~~~~ & ~~~~35 GeV~~~~ &~~~~20 GeV~~~~&~~~~4.0 GeV~~~~& ~~~~15 GeV~~~~& ~~~~$0.7\pi$~~~~& ~~~~280 GeV $\sim$ 325 GeV~~~~ \\
\hline ~~~~$m_\varphi =$ 320 GeV~~~~ & ~~~~40 GeV~~~~ &~~~~20 GeV~~~~&~~~~4.0 GeV~~~~& ~~~~15 GeV~~~~& ~~~~$0.7\pi$~~~~& ~~~~300 GeV $\sim$ 345 GeV~~~~ \\
\hline ~~~~$m_\varphi =$ 340 GeV~~~~ & ~~~~45 GeV~~~~ &~~~~20 GeV~~~~&~~~~4.0 GeV~~~~& ~~~~15 GeV~~~~& ~~~~$0.8\pi$~~~~& ~~~~320 GeV $\sim$ 365 GeV~~~~ \\
\hline ~~~~$m_\varphi =$ 360 GeV~~~~ & ~~~~45 GeV~~~~ &~~~~20 GeV~~~~&~~~~5.5 GeV~~~~& ~~~~15 GeV~~~~& ~~~~$\pi$   ~~~~& ~~~~340 GeV $\sim$ 385 GeV~~~~ \\
\hline ~~~~$m_\varphi =$ 380 GeV~~~~ & ~~~~40 GeV~~~~ &~~~~20 GeV~~~~&~~~~5.0 GeV~~~~& ~~~~15 GeV~~~~& ~~~~$\pi$   ~~~~& ~~~~355 GeV $\sim$ 405 GeV~~~~ \\
\hline ~~~~$m_\varphi =$ 400 GeV~~~~ & ~~~~50 GeV~~~~ &~~~~25 GeV~~~~&~~~~5.5 GeV~~~~& ~~~~20 GeV~~~~& ~~~~$\pi$   ~~~~& ~~~~370 GeV $\sim$ 435 GeV~~~~\\
\hline ~~~~$m_\varphi =$ 450 GeV~~~~ & ~~~~50 GeV~~~~ &~~~~25 GeV~~~~&~~~~6.5 GeV~~~~& ~~~~20 GeV~~~~& ~~~~$\pi$   ~~~~& ~~~~420 GeV $\sim$ 485 GeV~~~~\\
\hline ~~~~$m_\varphi =$ 500 GeV~~~~ & ~~~~45 GeV~~~~ &~~~~20 GeV~~~~&~~~~7.5 GeV~~~~& ~~~~20 GeV~~~~& ~~~~$\pi$   ~~~~& ~~~~465 GeV $\sim$ 535 GeV~~~~\\
\hline ~~~~$m_\varphi =$ 550 GeV~~~~ & ~~~~45 GeV~~~~ &~~~~20 GeV~~~~&~~~~8.0 GeV~~~~& ~~~~20 GeV~~~~& ~~~~$\pi$   ~~~~& ~~~~510 GeV $\sim$ 585 GeV~~~~\\
\hline ~~~~$m_\varphi =$ 600 GeV~~~~ & ~~~~50 GeV~~~~ &~~~~25 GeV~~~~&~~~~8.0 GeV~~~~& ~~~~25 GeV~~~~& ~~~~$\pi$   ~~~~& ~~~~555 GeV $\sim$ 640 GeV~~~~\\
\hline
\end{tabular}
\end{center}
\label{tab:cuts_ex}
\end{table*}%
\begin{table*}[htdp]
\caption{The combination of cuts which gives the largest significance for discovering the flavon with different mass.}
\begin{center}
\begin{tabular}{|c|c|c|c|c|c|c|}
\hline Discovery & $p_{T,{\text{cut}}}^{\gamma_1}$ &$p_{T,{\text{cut}}}^{\gamma_2}$ &$\Delta m_{h,{\text{cut}}}^{\gamma\gamma}$&$\Delta m_{h,{\text{cut}}}^{bb}$&$\Delta\phi_{\gamma b}$ & $m_{hh}$ window\\\hline
\hline ~~~~$m_\varphi =$ 260 GeV~~~~ & ~~~~30 GeV~~~~ &~~~~20 GeV~~~~&~~~~3.5 GeV~~~~&~~~~20 GeV~~~~& ~~~~$0.5\pi$~~~~ & ~~~~230 GeV $\sim$ 280 GeV~~~~ \\
\hline ~~~~$m_\varphi =$ 270 GeV~~~~ & ~~~~35 GeV~~~~ &~~~~20 GeV~~~~&~~~~3.5 GeV~~~~&~~~~20 GeV~~~~& ~~~~$0.5\pi$~~~~ & ~~~~245 GeV $\sim$ 295 GeV~~~~ \\
\hline ~~~~$m_\varphi =$ 280 GeV~~~~ & ~~~~30 GeV~~~~ &~~~~20 GeV~~~~&~~~~3.5 GeV~~~~&~~~~15 GeV~~~~& ~~~~$\pi$   ~~~~ & ~~~~265 GeV $\sim$ 300 GeV~~~~ \\
\hline ~~~~$m_\varphi =$ 290 GeV~~~~ & ~~~~30 GeV~~~~ &~~~~20 GeV~~~~&~~~~3.5 GeV~~~~&~~~~15 GeV~~~~& ~~~~$\pi$   ~~~~ & ~~~~275 GeV $\sim$ 310 GeV~~~~ \\
\hline ~~~~$m_\varphi =$ 300 GeV~~~~ & ~~~~35 GeV~~~~ &~~~~20 GeV~~~~&~~~~3.5 GeV~~~~&~~~~15 GeV~~~~& ~~~~$0.7\pi$~~~~ & ~~~~285 GeV $\sim$ 320 GeV~~~~ \\
\hline ~~~~$m_\varphi =$ 320 GeV~~~~ & ~~~~45 GeV~~~~ &~~~~20 GeV~~~~&~~~~4.0 GeV~~~~&~~~~15 GeV~~~~& ~~~~$0.7\pi$~~~~ & ~~~~300 GeV $\sim$ 345 GeV~~~~ \\
\hline ~~~~$m_\varphi =$ 340 GeV~~~~ & ~~~~45 GeV~~~~ &~~~~20 GeV~~~~&~~~~4.0 GeV~~~~&~~~~15 GeV~~~~& ~~~~$0.8\pi$~~~~ & ~~~~320 GeV $\sim$ 365 GeV~~~~ \\
\hline ~~~~$m_\varphi =$ 360 GeV~~~~ & ~~~~45 GeV~~~~ &~~~~25 GeV~~~~&~~~~5.5 GeV~~~~&~~~~15 GeV~~~~& ~~~~$0.8\pi$~~~~ & ~~~~340 GeV $\sim$ 385 GeV~~~~ \\
\hline ~~~~$m_\varphi =$ 380 GeV~~~~ & ~~~~50 GeV~~~~ &~~~~25 GeV~~~~&~~~~5.0 GeV~~~~&~~~~15 GeV~~~~& ~~~~$0.8\pi$~~~~ & ~~~~360 GeV $\sim$ 405 GeV~~~~ \\
\hline ~~~~$m_\varphi =$ 400 GeV~~~~ & ~~~~50 GeV~~~~ &~~~~25 GeV~~~~&~~~~5.5 GeV~~~~&~~~~15 GeV~~~~& ~~~~$\pi$   ~~~~ & ~~~~380 GeV $\sim$ 425 GeV~~~~ \\
\hline ~~~~$m_\varphi =$ 450 GeV~~~~ & ~~~~50 GeV~~~~ &~~~~25 GeV~~~~&~~~~6.5 GeV~~~~&~~~~15 GeV~~~~& ~~~~$\pi$   ~~~~ & ~~~~425 GeV $\sim$ 475 GeV~~~~ \\
\hline ~~~~$m_\varphi =$ 500 GeV~~~~ & ~~~~45 GeV~~~~ &~~~~20 GeV~~~~&~~~~6.5 GeV~~~~&~~~~15 GeV~~~~& ~~~~$\pi$   ~~~~ & ~~~~475 GeV $\sim$ 530 GeV~~~~ \\
\hline ~~~~$m_\varphi =$ 550 GeV~~~~ & ~~~~45 GeV~~~~ &~~~~25 GeV~~~~&~~~~7.5 GeV~~~~&~~~~15 GeV~~~~& ~~~~$\pi$   ~~~~ & ~~~~515 GeV $\sim$ 580 GeV~~~~ \\
\hline ~~~~$m_\varphi =$ 600 GeV~~~~ & ~~~~25 GeV~~~~ &~~~~25 GeV~~~~&~~~~7.0 GeV~~~~&~~~~15 GeV~~~~& ~~~~$\pi$   ~~~~ & ~~~~565 GeV $\sim$ 640 GeV~~~~ \\
\hline
\end{tabular}
\end{center}
\label{tab:cuts_dis}
\end{table*}
In FIG. \ref{fig:s300}, we show the results for the reconstructed diphoton, 
$b\bar b$, and $hh$ mass distributions using events which satisfy all the cuts.
The resonance signal is very clear.
The dominant background is $b\bar b\gamma\gamma$,  
and the other backgrounds are numerically small. The diphoton 
resonance is extremely clear,  and the $b\bar b$ peak is wide 
owing to the larger energy smearing of jets.  

After a scan over the mass of the flavon, we show expected limits from the search 
for the $\varphi \to hh\to b\bar b\gamma\gamma$ signal at 14 TeV with 
100 fb$^{-1}$ in FIG. \ref{fig:ex14hh}. The number of events is small 
so we calculate the $2\sigma$ exclusion bound using
\be
\sqrt{-2\left[n_b\ln\left(\frac{n_s+n_b}{n_b}\right)-n_s\right]}=2,
\ee
where $n_s$ and $n_b$ are the numbers of signal and background events, 
respectively \cite{Cowan:2010js}.  
The 5$\sigma$ discovery significance is calculated using 
\be
\sqrt{-2\left[\left(n_b+n_s\right)\ln\left(\frac{n_b}{n_s+n_b}\right)+n_s\right]}=5 .
\ee 
\begin{figure}[!htb]
\includegraphics[scale=0.35,clip]{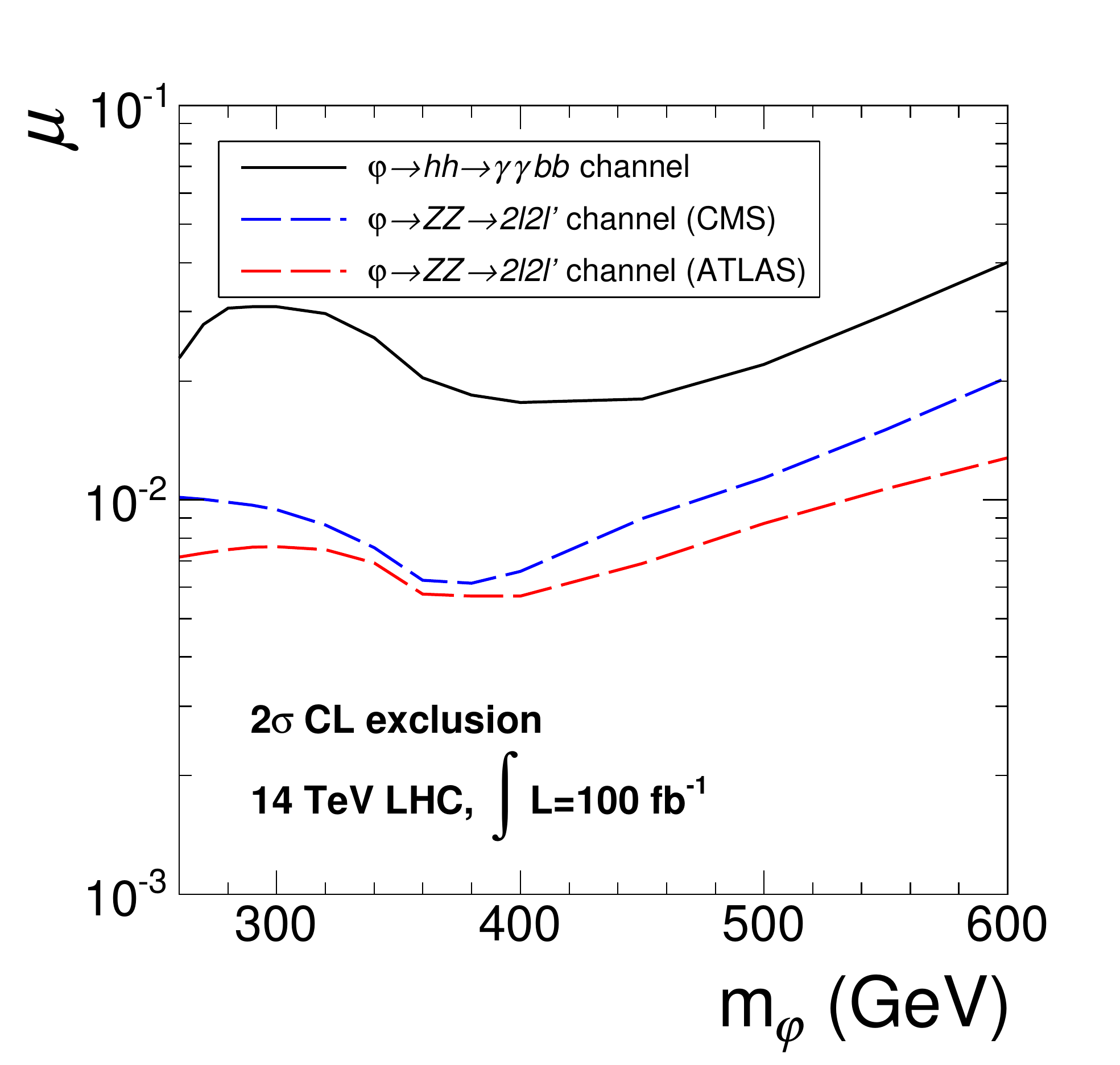}\\
\includegraphics[scale=0.35,clip]{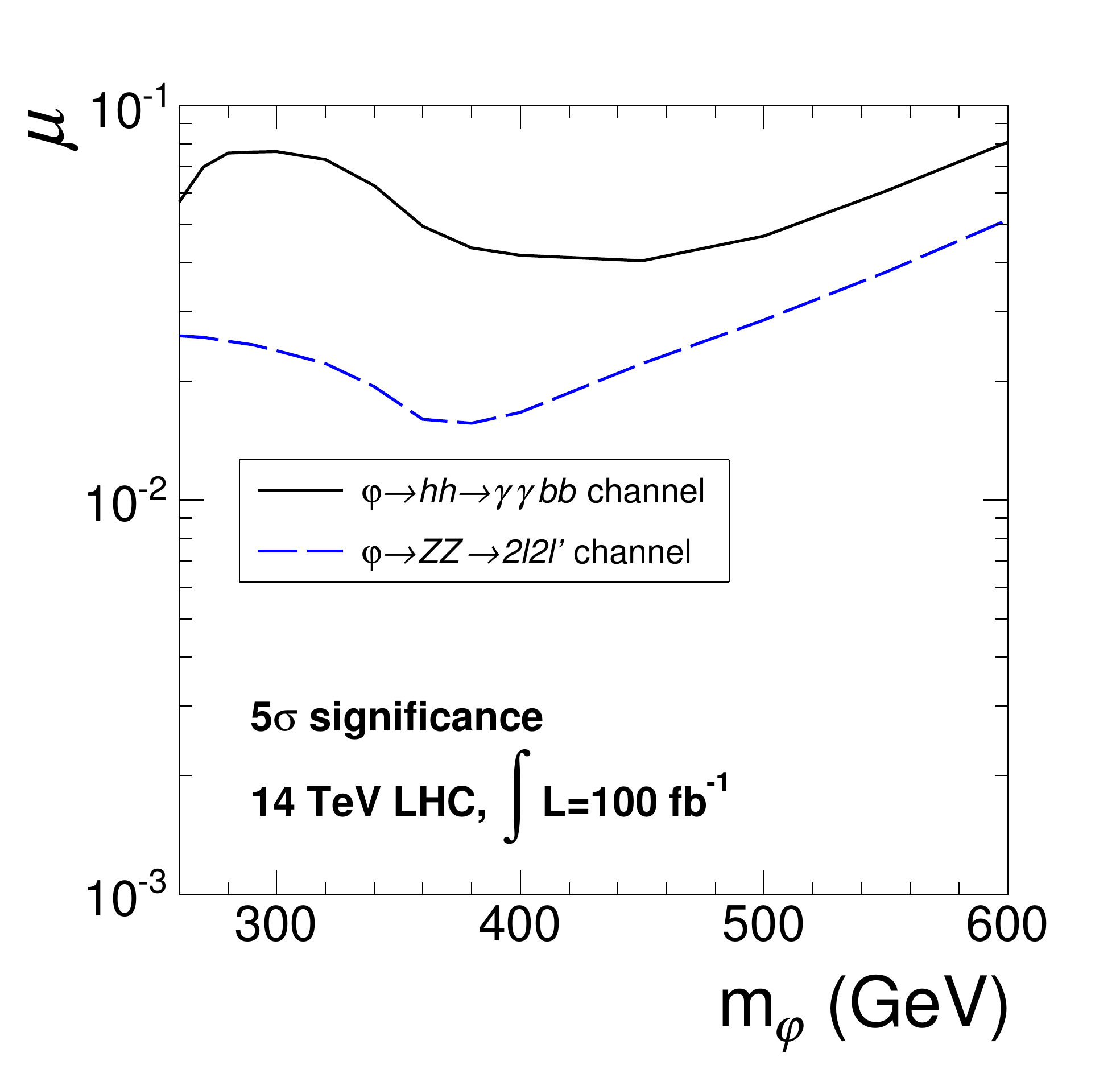}\\
\caption{The $2\sigma$ C.L. exclusion bound and the $5\sigma$ C.L. observation bound of the 
flavon production strength $\mu$ from the $pp\to \varphi \to hhX\to b\bar b \gamma\gamma X$ and the 
$pp \to\varphi  \to ZZX\to 2\ell2\ell^\prime$ channels. $\mu\equiv \sigma\left(pp\to \varphi \to ZZ\right)/
\sigma\left(pp\to H\right)$ for the $ZZ$ channel and $\mu\equiv \sigma\left(pp\to \varphi \to hh\right)/
\sigma\left(pp\to H\right)$ for the $hh$ channel. The region above the curves
will be excluded (discovered) at $2\sigma$ ($5\sigma$) C.L..
}
\label{fig:ex14hh} 
\end{figure}
The limits shown in FIG. \ref{fig:ex14hh} can be translated into a constraint on 
the NP parameters.  \begin{figure}[!htb]
\includegraphics[scale=0.35,clip]{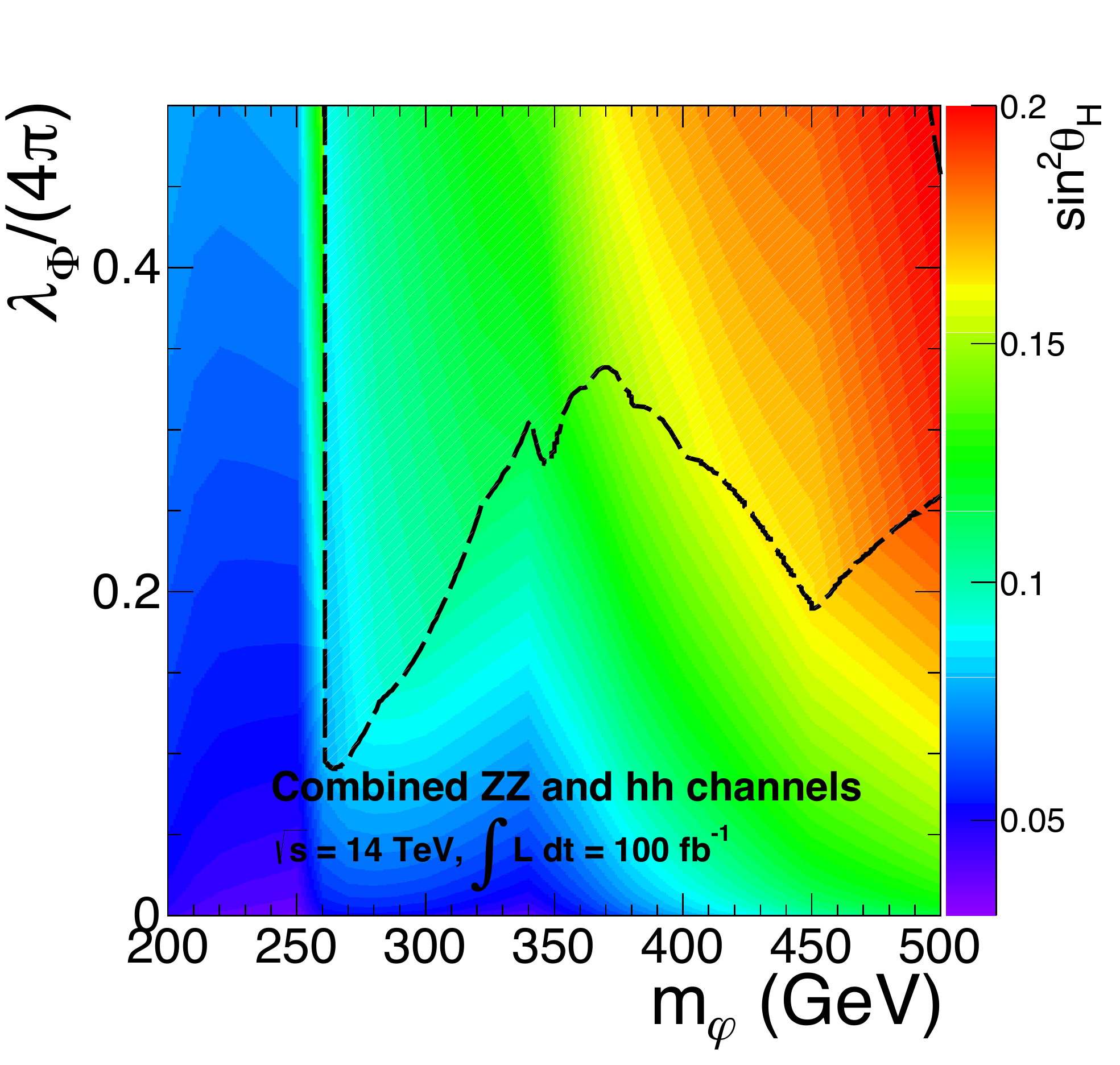}
\caption{Combination of the 
2$\sigma$ exclusion region of $s_H^2$
results for the $\varphi \to Z^0Z^0 \to 2\ell2\ell^\prime$ search
and the $\varphi \to hh \to b\bar b\gamma
\gamma$ search.  In the upper part of the figure, in the irregularly 
shaped region above the broad-dashed line,  
the $\varphi \to hh \to b \bar b \gamma \gamma$ 
search yields a stronger constraint.
\label{fig:14hh} }
\end{figure}
\begin{figure}[!htb]
\includegraphics[scale=0.35,clip]{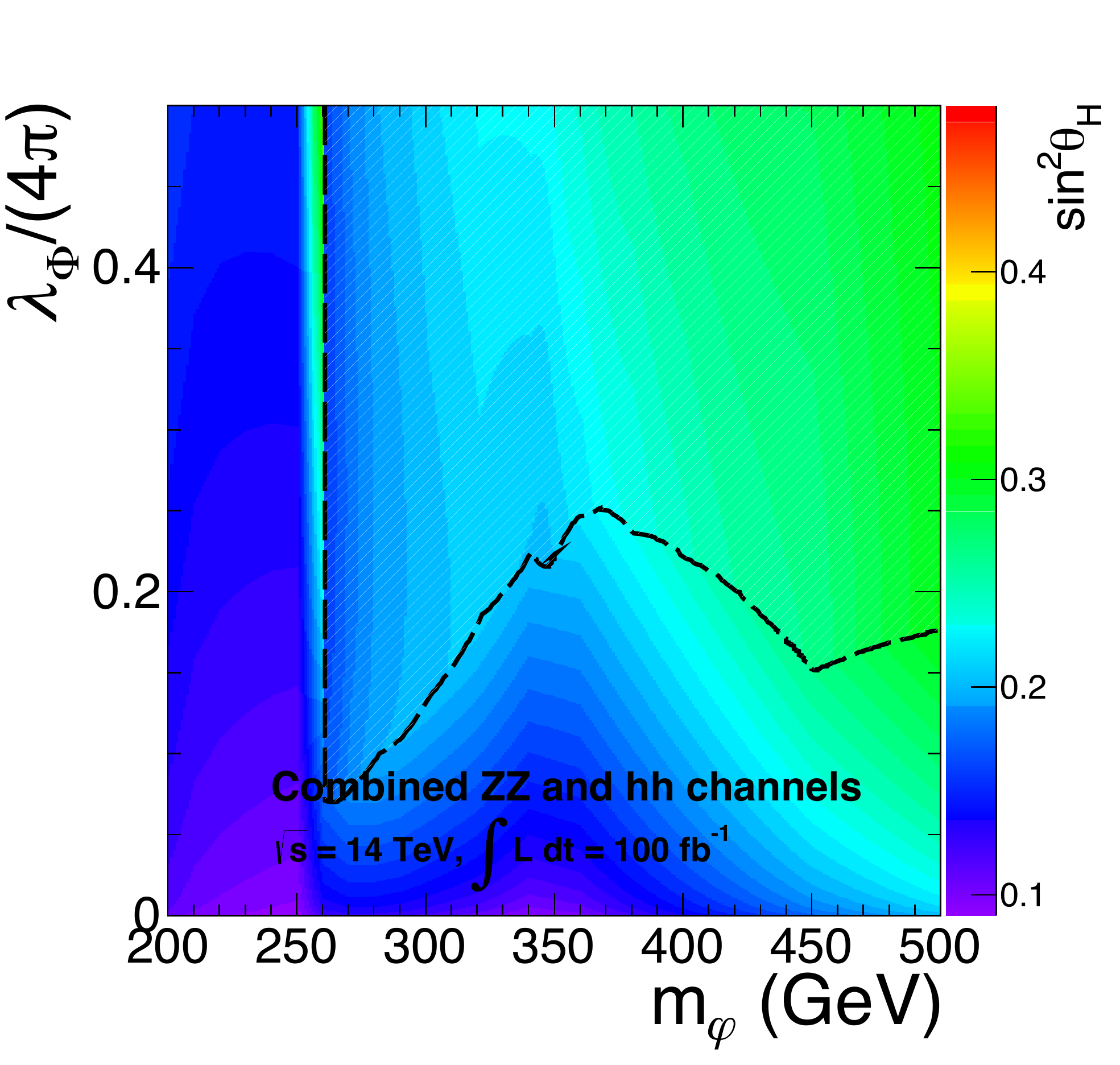}
\caption{The 5$\sigma$ discovery significance of the required value of $s_H^2$ 
from a combination of the $\varphi \to Z^0Z^0 \to 2\ell2\ell^\prime$ search
and the $\varphi \to hh \to b\bar b\gamma
\gamma$ search.  In the upper part of the figure, in the irregularly 
shaped region above the broad-dashed line,  
the $\varphi \to hh \to b \bar b \gamma \gamma$ process is more sensitive to the NP model.}
\label{fig:14hh5} 
\end{figure}

Combining the results from the SM-like heavy Higgs boson search 
for $\varphi \rightarrow Z^0Z^0$ and the $\varphi \to hh\to b\bar b \gamma\gamma$ search 
at 14 TeV with 100 fb$^{-1}$ integrated luminosity,
we show the constraint on $s_H^2$ in FIG. 
\ref{fig:14hh}.  The combined 5$\sigma$ discovery 
significance is shown in FIG. \ref{fig:14hh5}. 
In FIG. \ref{fig:14hh}, the search for the 
$\varphi \to hh\to b\bar b \gamma\gamma$ signal gives a stronger constraint 
in the cross-hatched region.  This signal can give 
a stronger constraint in the large $\lambda_\Phi$ 
region when the $\varphi\to hh$ channel opens.
This figure shows that a strong constraint on 
the neutral scalar $\varphi$ can be obtained with a combination of the two channels.

In this section, we investigated flavon phenomenology at the LHC
in detail.  We first examined the decay branching ratios and the production 
cross section of the flavon $\varphi$.  The dominant decay channels are
$W^+W^-$, $Z^0Z^0$, $hh$ and $t\bar t$.  We checked the limits on the 
flavon from heavy Higgs boson searches at 7 and 8 TeV.   The 
$hh$ signal was simulated in detail at 14 TeV.  We showed that in some parts of 
parameter space the search for a $hh$ signal yields a stronger constraint on 
the NP model than the $Z^0Z^0$ channel.  Finally, we presented a combined
result using the $hh\to2b2\gamma$ and the $Z^0Z^0\to2\ell2\ell^\prime$ channels.

\section{Conclusions}
\label{sec:con}
This paper has investigated a model of physics beyond the SM in 
which there is a new scalar, a flavon,  a new heavy fermion associated with 
the SM top-quark, and a new neutral flavor gauge boson. 
This model arises as the low-energy limit of a theory of gauged 
flavor symmetry with an inverted hierarchy, giving a simplified 
model with spontaneous breaking of flavor symmetry.  This model may be considered independently 
of its origin from the inverted hierarchy, as a model that could 
result from other physics and be interesting in its own right.
The flavon mixes with the SM 
Higgs boson, and the heavy fermion alters the production and decay properties of 
the Higgs boson at the LHC, all in ways that are consistent with data at current 
levels of precision. The flavon and the heavy fermion might appear at the hundreds of 
GeV to the TeV scale. There is a sizable allowed parameter space in which existing constraints
from electroweak precision observables and flavor physics are satisfied.

The mixing of the flavon, which is a SM gauge singlet scalar, with the $SU(2)_L$ doublet Higgs
field produces two notable effects.  First, its influence on Higgs boson physics could
be examined with more precise measurements of the SM Higgs-like scalar at 
125 GeV discovered recently at the LHC. Second, the mixing makes it possible to produce 
and detect the flavon at the LHC.

In this NP model, the production cross section of the SM-like Higgs boson at the LHC
is suppressed by a factor $\cos^2 \theta_H$, where $\theta_H$ is the mixing angle between 
the Higgs boson and the flavon.  However, neither mixing nor the triangle loop from the heavy 
fermion change the Higgs boson decay branching ratios significantly.  The $h\to Z^0\gamma$ decay 
channel is an exception. With large mixing which is still allowed at the 3$\sigma$ level,
the branching ratio of this channel can be increased by about 5\%. It is not an easy
task to measure this branching ratio precisely at the LHC, but it would be possible to
check the modification of the $hZ^0\gamma$ vertex at a future Higgs Factory.

The possibility to search for the flavon at the LHC is explored in detail in the paper.
Generally, the mixing between an exotic scalar field $\varphi$ and the SM Higgs field can be
generated from $\mathcal{O}\left(\Phi\right)\left(H^\dagger H\right)$, where $\mathcal{O}
\left(\Phi\right)$ is some operator constructed from $\Phi$. As long as the scalar
interacts with the SM sector, this interaction will arise from loop corrections even if
forbidden artificially at tree-level. The $\varphi hh$ vertex usually appears once there is
$\varphi-h$ mixing, and a NP heavy scalar boson which can decay into a SM Higgs-pair 
will also decay via the SM Higgs boson decay modes. This is also the case in this
flavor symmetry model where the flavon decays to a SM Higgs-pair, $\varphi \rightarrow hh$,
as well as through the SM Higgs boson decay modes. 

We investigated the $\varphi \rightarrow Z^0Z^0$ discovery channel.  At 7 and 8 TeV at the
LHC, the $Z^0Z^0$ channel will give a stronger constraint than $\varphi \rightarrow h h$
owing to limitations of integrated luminosity. The large mixing required to get a
large enough cross section is excluded by the global-fit of the Higgs 
boson inclusive cross section. At 14 TeV with 100 fb$^{-1}$ integrated luminosity, we showed that 
the small mixing region can be reached where the $Z^0Z^0$ decay channel is highly 
suppressed. In this region of the parameter space, the $\varphi \rightarrow hh$ signal is more 
important for discovery. Our result can be used to discover or exclude not only the flavon,
but also singlet scalars in other models which can be produced through gluon fusion  
and decay into a Higgs pair \cite{O'Connell:2006wi,Barger:2007im,Xiao:2014kba,He:2014ora}.

The SM Higgs pair production cross section is changed in this model of NP.   The flavon 
can be produced singly at the LHC.  If it decays into the $hh$ final state with a sizable decay 
branching ratio, the $hh$ cross section will be enhanced significantly by this resonance effect.  
A second source of change comes from corrections to the $h\bar tt$ and $hhh$ vertices.   
These corrections generally reduce the SM Higgs pair rate as shown in FIG. \ref{fig:hhsm}.

In this work, we investigate the case $m_T>m_\varphi $ in which the new heavy fermion is more
massive than the flavon. Such a heavy fermion has special decay modes and signals at the
LHC. A relatively heavy flavon and the new massive vector boson $Z_{T\mu}$ might 
produce a four-top signal at the LHC, which is a potential discovery channel for these states.
A study of the heavy fermion phenomenology and the top-philic vector boson phenomenology
is left for another paper \cite{Haozhang:2014pr}.

\begin{acknowledgments}
We thank Carlos E.M. Wagner and Ian Low for helpful discussions. 
The work of E. L. Berger at Argonne is supported 
in part by the U.S. DOE under Contract No. DE-AC02-06CH11357.
S. Giddings and H. Zhang were supported by the U.S. DOE under Contract No. 
DE-FG02-91ER40618. S. Giddings is also supported in part by 
Foundational Questions Institute grant number FQXi-RFP3-1330.
H. Wang is supported in part by the U.S. DOE
under DOE Contract DE-AC02-05CH11231.  E.~L.~Berger warmly acknowledges 
the hospitality of the Kavli Institute for Theoretical Physics , University of California, Santa Barbara 
where his research was 
supported in part by the National Science Foundation under Grant No. NSF PHY11-25915. 
H. Zhang warmly acknowledges 
the hospitality of Argonne National Laboratory
where his research was 
supported in part by the U.S. DOE under Contract No. DE-AC02-06CH11357, 
and the Center for High Energy Physics at Peking University, and 
the Department of Physics and Astronomy at Shanghai Jiao Tong University.
\end{acknowledgments}  

\bibliographystyle{apsrev}
\bibliography{draft}
  
\end{document}